\documentclass[superscriptaddress,twocolumn,english,floatfix,aps,prb,amsmath,amssymb,longbibliography]{revtex4-1}
\usepackage[T1]{fontenc}
\usepackage[latin9]{inputenc}
\usepackage{color}
\usepackage{times}
\usepackage{babel}

\usepackage{amsmath}
\usepackage{amssymb}
\usepackage{graphicx}
\usepackage{esint}
\usepackage[colorlinks=true]{hyperref}
\hypersetup{citecolor=blue,linkcolor=blue,urlcolor=blue}
\begin{document}

\title{Two-dimensional materials in the presence of nonplanar interfaces} 

\author{Danilo T. Alves}
\email{danilo@ufpa.br}
\affiliation{Faculdade de F\'{i}sica, Universidade Federal do Par\'{a}, 66075-110, Bel\'{e}m, Par\'{a}, Brazil}
\affiliation{Centro de F\'{i}sica, Universidade do Minho, P-4710-057, Braga, Portugal}

\author{N. M. R. Peres}
\email{peres1975@gmail.com}
\affiliation{Centro de F\'{i}sica, Universidade do Minho, P-4710-057, Braga, Portugal}
\affiliation{Departamento de F\'{i}sica, Universidade do Minho, P-4710-057, Braga, Portugal}
\affiliation{International Iberian Nanotechnology Laboratory (INL), Av. Mestre Jos\'{e} Veiga, 4715-330, Braga, Portugal}

\date{\today}

\begin{abstract}
We consider a planar two-dimensional system 
between two media with different dielectric constants and in 
the presence of a third dielectric medium separated by a nonplanar interface.
Extending a perturbative method for solving Poisson's equation, developed by Clinton, Esrick, and Sacks 
[Phys. Rev. B, \textbf{31}, 7540 (1985)], in
the presence of nonplanar conducting boundaries to the situation proposed here, we obtain,
up to the first order in terms of the function which defines the nonplanar interface,
the effective potential, the effective electrostatic field, the effective dielectric constant for the planar 2D system,
and the effective external field acting in-plane in the 2D system. 
Implications of the results to properties of 2D systems are discussed.
In the limit of planar surfaces, vacuum-dielectric or vacuum-conducting media,
our results are in agreement with those found in the literature.
\end{abstract}

%\pacs{...}
%AJUSTAR PACS

\maketitle

%%%%%%%%%%%%%%%%%%%%%%%%%%%%%%%%%%%%%%%%%%%%%%%%%%%%%%%%%%%%%%%%%%%%%%%%%%%%%%%%%%%%%%%%%%%
\section{Introduction}

Two-dimensional (2D) systems, as for example a 2D elecron-gas in a heterostructure or in doped graphene,
have properties influenced by the electron-electron interaction
\cite{Jang-1993,Zheng-1993,Vozmediano-1994,Vozmediano-2012,Geim-2011-Nature,Santos-2013},
as well as by the presence of external electric and magnetic fields \cite{Nuno-Livro}.
In the case of graphene, the electron-electron interactions
implies in the renormalization of the Fermi velocity, thus
reshaping the Dirac cones \cite{Vozmediano-1994,Marino-2017-livro}, 
an effect that was experimentally observed \cite{Geim-2011-Nature}. 

In the context of quantum field theories applied to the condensed matter, the pseudo-quantum electrodynamics (PQED)
(sometimes called reduced quantum electrodynamics), an effective and complete description in 2+1 dimensions for electronic systems moving on a plane, was built considering that the static potential of interaction between electrons
in the 2D system should be Coulombian, instead of the logarithmic one ($\propto \ln r $) characteristic of quantum electrodynamics in 2+1 dimension
\cite{marino-1993,Marino-2017-livro}.
On the other hand, the effective interaction between electrons in a two-dimensional system  
can be changed by the presence of material media. For example, 
 it was recently shown that the logarithmic renormalization of the Fermi velocity 
in a plane graphene sheet (which, in turn, is related to the Coulombian static potential associated to electrons in the sheet) 
is inhibited by the presence of a single parallel plate or a cavity formed by conducting plates \cite{Grupo-NPB-I,Grupo-NPB-II},
with this inhibition leading to an increase of the optical conductivity. 

In addition,  the effective interaction between electric charges in a two-dimensional planar system,
when it is put in the presence of a planar interface between
dielectric media \cite{Profumo-2010,Katsnelson-2011,Nuno-Livro},   has been investigated. This change of the electron-electron interaction due to the presence of boundaries affects Coulomb drag between graphene single layers \cite{Katsnelson-2011,Polini-drag}.

The problem of finding the effective interaction between static charges in a two-dimensional system
can be viewed as part of a class of problems focusing on a static point charge in the presence of an interface between
two media. 

Essentially, the field of the charge induces an electric polarization
on the interface (or a surface charge distribution), which generates an additional electric field, usually named image field,
whose knowledge enables us to find the effective potential and external fields acting on the two-dimensional system
where the point charge lives. 

In the 1970s, the image potential was discussed in the context of several
phenomena. For instance, the image-potential states, which are quantum states of
electrons localized at surfaces of materials which exhibit
negative electron affinity \cite{Clinton-1985-I}. 
These electrons cannot escape from the surface due to
the image electric potential field and cannot penetrate into the material due to the negative electron
affinity \cite{Echenique-1991}, as it occurs with electrons in the vicinity of a liquid-helium interface
\cite{Cole-1970,Clinton-1985-I,Echenique-1991}.
On the other hand, up to 1980, the majority of cases that had been investigated 
of image-potential effects assumed that the interfaces between the media
were planar \cite{Maradudin-1980,Sun-1990}. 
Motivated by the fact that it is almost impossible to create a perfectly
planar surface and interested in
determining effects of corrugation on the image potential, 
Rahman and Maradudin \cite{Maradudin-1980} calculated perturbatively the electrostatic image potential for a point charge located near a rough vacuum-isotropic dielectric interface, with the surface of separation described by a random function with mean value equal to zero. 
The problem of finding the image potential for a point charge in vacuum in the presence of a
nonplanar metal surface has been investigated by Clinton \textit{et al.}\cite{Clinton-1985-I}, 
who, based on a work-energy argument, obtained a general formulation for the image potential for first-order deformations of an arbitrary shape,
showed that ions and electrons are always attracted to the elevated part of the surface \cite{Clinton-1985-I}.
Clinton \textit{et al.}\cite{Clinton-1985-II} also presented 
a formal solution for the electrostatic potential by solving perturbatively 
Poisson's equation in the presence 
of a generally modified planar conducting surface, with the solution extendable to any perturbation 
order in the corrugation function \cite{Clinton-1985-II}. 

Non planar interfaces occur naturally in graphene-based plasmonic systems
\cite{Koppens-Nuno-2018}. In this class of systems, patterned metallic gratings
are positioned at a distance of a single atom from a single graphene sheet, thus leading naturally to the class of problems discussed in this paper. Also, the problem of 
nanoparticles deposited on graphene, 
\cite{Bruno-Dyadic}
and how they change the electron-electron interactions in graphene, is another class of problems that can be solved using the 
approach we develop ahead. Naturally, the presence of metallic substrates near a 2D material changes the optical conductivity of the material. How this change occurs is also controlled by the nature of the interface near the material and, therefore, incorporating the effect of corrugation in the formalism is a natural application of the problems tackled in this work. 

Doped transition-metal dichalcogenides are known to have strong electron-hole interactions (excitonic effects) which can be tuned by the presence of interfaces, being they of dielectric or metallic nature \cite{Eduardo-Nonlocal}.
 Again, how the presence of corrugation changes the electron-electron interaction in this class of systems is a highly relevant problem in the field of 2D materials. Finally, if the corrugation 
\cite{Andre-Gil}
occurs in the scale of tens of nanometers, a length scale well in reach of microfabrication techniques, the corrugation plays the role of a scattering potential for the electronic propagation, thus affecting the DC conductivity of the electrons in the 2D material. Since hexagonal Boron Nitride has allowed an unprecedent control on the distance a 2D system can be positioned near a corrugated interface, the problem discussed in this paper acquires relevance
for applications in the field of polaritonics using 2D materials. 

In the present paper, we investigate how the presence of nonplanar surfaces changes the effective
electrostatic interaction between electrons in a two-dimensional system, 
producing an effective potential dependent not only on the distance to the source charge
but also on the position of the charge itself, and also how nonplanar surfaces generate an effective in-plane
external electric field acting along the two-dimensional system.
Specifically, considering a typical configuration \cite{Katsnelson-2011,Nuno-Livro},
we investigate a planar two-dimensional system 
between two media with different dielectric constants, in 
the presence third dielectric medium separated by a nonplanar interface.
Extending the perturbative method for solving Poisson's equation in
the presence of nonplanar conducting boundaries, proposed by Clinton, Esrick, and Sacks 
\cite{Clinton-1985-II}, to the situations discussed here, we obtain the first correction to
the effective potential and dielectric constants for the planar two-dimensional system,
as well as calculate the coordinate dependent external electric field
induced by the nonplanar surface.
As an application of our results, we use our results to the case of sinusoidal surfaces.
Finally, implications of the results to properties of two-dimensional systems are discussed.

The paper is organized as follows. In Sec. \ref{main-calculations} we obtain the total electric potential
function for the problem of a point charge 
between two media with different dielectric constants, and in 
the presence third dielectric medium separated by a nonplanar surface. We obtain, from our formulas, 
the particular results for two dielectrics,
vacuum-dielectric and vacuum-conducting media, extending and recovering results found in the
literature. We obtain the effective potential and dielectric constants 
for charges living in a 2D planar system put between two dielectric media,
also showing the appearance of an effective external field, induced by
the nonplanar interface, acting on the charges in this 2D system. 
In Sec. \ref{app-point-charge},
we apply our formulas to the case of sinusoidal surfaces and, using realistic values,
obtain estimates for the intensities of effective interaction and external field.
In Sec. \ref{comments-implications}, we present our final comments as well as discuss some implications
of our results for two-dimensional systems.

%%%%%%%%%%%%%%%%%%%%%%%%%%%%%%%%%%%%%%%%%%%%%%%%%%%%%%%%%%%%%%%%%%%%%%%%%%%%%%%%%%%%%%%%%%%%%%%%
\section{Point-charges confined between two dielectrics, in the presence
of a third dielectric region with a nonplanar interface}
\label{main-calculations}

%%%%%%%%%%%%%%%%%%%%%%%%%%%%%%%%%%%%%%%%%%%%%%%%%%%%%%%%%%%%%%%
\subsection{Statement of the problem} 

\begin{figure}[t]
\begin{centering}
\includegraphics[width=0.5\columnwidth]{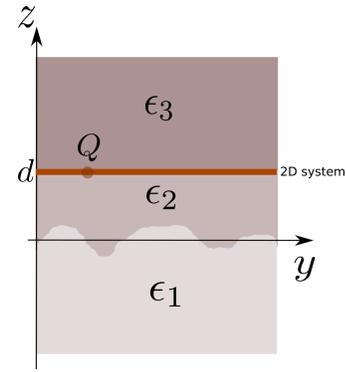} 
\end{centering}
\caption{Illustration of the configuration formed by three different dielectric media
($x$-axis perpendicular to the paper).
One can see a planar interface between the regions $\epsilon_3$ and $\epsilon_2$, 
where a two-dimensional system is located. All charges of this system, for instance the charge $Q$ illustrated in the figure, 
are confined to this plane.
The figure also shows a nonplanar interface [described by $z=\lambda h\left({\bf r}_{||}\right)$] separating the regions $\epsilon_2$ and $\epsilon_1$.}
\label{3-camadas-1-rugosa-carga} 
\end{figure}

We  consider a stratified  medium containing three different insulators , arranged as
as in Fig. \ref{3-camadas-1-rugosa-carga} (for the purposes of this paper, the first dielectric can also be replaced by a metallic medium). Mathematically, the position of the dielectrics 
are given by:
\begin{equation}
 \epsilon\left({\bf r}\right) = 
  \begin{cases} 
   \epsilon_3,\; z>d \\
	 \epsilon_2,\; \lambda h\left({\bf r}_{||}\right)<z<d \\
   \epsilon_1,\; z<\lambda h\left({\bf r}_{||}\right),
  \end{cases}
	\label{epsilon-com-chaves}
\end{equation}
where  ${\bf r}_{||}=x\hat{{\bf x}}+y\hat{{\bf y}}$, 
$d>0$, $z=\lambda h({\bf r}_{||})$ [$h({\bf r}_{||})<d$]  defines a general (nonplanar) surface,
and $\lambda$ ($|\lambda|<1$) is a dimensionless parameter such that for $\lambda=0$
one recovers the planar surface case at $z=0$. We consider $\hat{{\bf x}}$, $\hat{{\bf y}}$
and $\hat{{\bf z}}$ unit vectors pointing to the $x$, $y$, and $z$ directions, respectively.
For practical purposes, we write
\begin{align}
\epsilon\left({\bf r}\right)=&\epsilon_{3}\theta\left(z-d\right)+\epsilon_{2}\theta\left[z-\lambda h\left({\bf r}_{||}\right)\right]
\theta\left(d-z\right)
\nonumber\\
&+\epsilon_{1}\theta\left\{-\left[z-\lambda h\left({\bf r}_{||}\right)\right]\right\},
\label{epsilon-practical}
\end{align}
whose expansion in $\lambda$ leads to
\begin{eqnarray}
\epsilon\left({\bf r}\right)&=&\epsilon_{3}\theta\left(z-d\right)
+\epsilon_{2}\theta\left(z\right)\theta\left(d-z\right)+\epsilon_{1}\theta\left(-z\right)
\cr
&+&
\delta\left(z\right)\left[-\epsilon_{2}\theta\left(d-z\right)+\epsilon_{1}\right]\lambda h({\bf r}_{||})
+\mathcal{O}(\lambda^2)+...
\label{epsilon-expanded}
\end{eqnarray}

We consider the problem of a two-dimensional system of point charges confined in the
planar interface between the media $\epsilon_3$ and $\epsilon_2$, as illustrated in Fig. \ref{3-camadas-1-rugosa-carga}. This is achieved, positioning a 2D system between these two dielectrics.
From Gauss's law, we have, for a charge $Q$ located at the position ${\bf r}={\bf r}^{\prime}$,
with ${\bf r}^{\prime}={\bf r}_{||}^{\prime}+d\hat{{\bf z}}$,
\begin{equation}
\boldsymbol{\nabla}\cdot\left[\epsilon\left({\bf r}\right)\boldsymbol{\nabla}{\bf \phi\left({\bf r},{\bf r}^{\prime}\right)}\right]=-4\pi Q\delta\left({\bf r}-{\bf r}^{\prime}\right),
\label{gauss-law}
\end{equation}
where the potential $\phi$ can be written as 
$\phi\left({\bf r}, {\bf r}^{\prime}\right)=QG\left({\bf r},{\bf r}^{\prime}\right)$.
Following Clinton, Esrick, and Sacks \cite{Clinton-1985-II}, we look for a solution
of $G$ as an expansion in powers of $\lambda$:
\begin{equation}
G\left({\bf r},{\bf r^{\prime}}\right)=G^{(0)}\left({\bf r},{\bf r^{\prime}}\right)+\sum_{n=1}^{\infty}\lambda^{n}G^{(n)}\left({\bf r},{\bf r^{\prime}}\right),
\label{G-expanded}
\end{equation}
where $G^{(0)}\left({\bf r},{\bf r^{\prime}}\right)$ is related to the solution
of Gauss's equation for planar interfaces.
To solve Eq. (\ref{gauss-law}) with $\epsilon\left({\bf r}\right)$ given by Eq. (\ref{epsilon-practical}), 
it is convenient to introduce the Fourier transform in the $x$, $y$ coordinates,
\begin{equation}
f({\bf r}_{||},z)=\int\frac{1}{\left(2\pi\right)^{2}}d^{2}{\bf q}f\left({\bf q},z\right)e^{i{\bf q}\cdot{\bf r}_{||}},
\label{fourier-representation}
\end{equation}
where $f$ can represent any function of ${\bf r}_{||}$ considered in the present paper,
and ${\bf q}=q_x\hat{{\bf x}}+q_y\hat{{\bf y}}$. We also have
\begin{equation}
f\left({\bf q},z\right)=\int d^{2}{\bf r}f\left({\bf r}_{||},z\right)e^{-i{\bf q}\cdot{\bf r}_{||}},
\label{fourier-inverse-representation}
\end{equation}
and we are adopting the same nomenclature for a given function of $({\bf r}_{||},z)$ and
for its 2D Fourier transform.
Using the representation given in Eq. (\ref{fourier-representation}), it can be shown that Eq. (\ref{gauss-law}) can be written as
\begin{multline}
\int\frac{1}{\left(2\pi\right)^{2}}d^{2}{\bf q^{\prime}}\bigg\{ \left({\bf q^{\prime}}\cdot{\bf q}\right)G\left({\bf q^{\prime}},z,{\bf r}^{\prime}\right)\epsilon\left({\bf q}-{\bf q^{\prime}},z\right)
\\
-\frac{\partial}{\partial z}\left[\frac{\partial G\left({\bf q^{\prime}},z,{\bf r}^{\prime}\right)}{\partial z}\epsilon\left({\bf q}-{\bf q^{\prime}},z\right)\right]\bigg\} 
\\
=4\pi\delta\left(z-d\right)e^{-i{\bf q}\cdot{\bf r}_{||}^{\prime}}.
\label{gauss-law-fourier}
\end{multline}
The continuity condition for $G\left({\bf q},z,{\bf r}^{\prime}\right)$
is required for all values of $z$.
Specifically focusing on the interfaces, we have ($\eta>0$):
\begin{equation}
\lim_{\eta\rightarrow 0}G\left({\bf q},d+\eta,{\bf r}^{\prime}\right)
=\lim_{\eta\rightarrow 0}G\left({\bf q},d-\eta,{\bf r}^{\prime}\right),
\label{bc-main-1}
\end{equation}
\begin{equation}
\lim_{\eta\rightarrow 0}G\left({\bf q},\lambda h\left({\bf r}_{||}\right)+\eta,{\bf r}^{\prime}\right)
=\lim_{\eta\rightarrow 0}G\left({\bf q},\lambda h\left({\bf r}_{||}\right)-\eta,{\bf r}^{\prime}\right).
\label{bc-main-2}
\end{equation}
Note that these boundary conditions apply to the full Green's function.
%%%%%%%%%%%%%%%%%%%%%%%%%%%%%%%%%%%%%%%%%%%%%%%%%%%%%%%%%%%%%%%
\subsection{Method of solution} 

The central point of the present calculation is to substitute 
(\ref{epsilon-expanded}) and (\ref{G-expanded}) into (\ref{gauss-law-fourier}),
and requiring that the coefficients of $\lambda^{n}$ vanish. This yields 
(up to first order in $\lambda$)  an equation for $G^{(0)}$, 
\begin{multline}
\boldsymbol{\nabla}\cdot\bigg\{\bigg[\epsilon_{3}\theta\left(z-d\right)+\epsilon_{2}\theta\left(z\right)\theta\left(d-z\right)+\epsilon_{1}\theta\left(-z\right)\bigg]
\\
\boldsymbol{\nabla}\left[G^{(0)}\left({\bf r},{\bf r^{\prime}}\right)\right]\bigg\}
 =-4\pi\delta\left({\bf r}-{\bf r}^{\prime}\right),
\label{G0-equation} 
\end{multline}
and an equation for $G^{(1)}$,
\begin{multline}
\boldsymbol{\nabla}\cdot\bigg\{ \epsilon_{3}\theta\left(z-d\right)+\epsilon_{2}\theta\left[z\right]\theta\left(d-z\right)
+\epsilon_{1}\theta\left(-z\right)
\\
\boldsymbol{\nabla}\left[G^{(1)}\left({\bf r},{\bf r^{\prime}}\right)\right]\bigg\}=\epsilon_{21}^{-}\boldsymbol{\nabla}\cdot\left\{ \delta\left(z\right)h\left({\bf r}_{||}\right)
\boldsymbol{\nabla}\left[G^{(0)}\left({\bf r},{\bf r^{\prime}}\right)\right]\right\}, 
\label{G1-equation}
\end{multline}
where hereafter we consider
\begin{equation}
\epsilon_{ij}^{\pm}=\epsilon_{i}\pm\epsilon_{j}.
\end{equation}
These equations in Fourier space are given, respectively, by:
\begin{multline}
q^{2}\epsilon\left(z\right)G^{(0)}\left({\bf q},z,{\bf r}_{||}^{\prime},d\right)-\frac{\partial}{\partial z}\left[\epsilon\left(z\right)\frac{\partial G^{(0)}\left({\bf q},z,{\bf r}_{||}^{\prime},d\right)}{\partial z}\right]
\\
=4\pi\delta\left(z-d\right)e^{-i{\bf q}\cdot{\bf r}_{||}^{\prime}}
\label{G0-equation-fourier} 
\end{multline}
and
\begin{multline}
q^{2}G^{(1)}\left({\bf q},z,{\bf r}^{\prime}\right)\bigg[\epsilon_{3}\theta\left(z-d\right)+\epsilon_{2}\theta\left(z\right)\theta
\left(d-z\right)+\epsilon_{1}\theta\left(-z\right)\bigg]
\\
-\frac{\partial}{\partial z}\bigg\{ \frac{\partial G^{(1)}\left({\bf q},z,{\bf r}^{\prime}\right)}{\partial z}
\\
\bigg[\epsilon_{3}\theta\left(z-d\right)
+\epsilon_{2}\theta\left(z\right)\theta\left(d-z\right)+\epsilon_{1}\theta\left(-z\right)\bigg]\bigg\} 
\\
=\epsilon_{21}^{-}\int\frac{1}{\left(2\pi\right)^{2}}d^{2}{\bf q^{\prime}}\bigg\{ \left({\bf q^{\prime}}\cdot{\bf q}\right)G^{(0)}\left({\bf q^{\prime}},z,{\bf r^{\prime}}\right)
\\
\delta\left(z\right)h\left({\bf q}-{\bf q^{\prime}}\right)
-\frac{\partial}{\partial z}\left[\frac{\partial G^{(0)}\left({\bf q^{\prime}},z,{\bf r^{\prime}}\right)}{\partial z}\delta
\left(z\right)h\left({\bf q}-{\bf q^{\prime}}\right)\right]\bigg\}. 
\label{G1-equation-fourier}
\end{multline}

The solution for (\ref{G0-equation-fourier}), taking into account that
$G^{\left(0\right)}$ is continuous through the interfaces, is known and given by \cite{Nuno-Livro,Katsnelson-2011,Profumo-2010}:
\begin{equation}
 G^{\left(0\right)}\left({\bf q},z,{\bf r}^{\prime}\right) = 
  \begin{cases} 
   G_{III}^{\left(0\right)}\left({\bf q},z,{\bf r}_{||}^{\prime},d\right),\; z\geq d \\
   G_{II}^{\left(0\right)}\left({\bf q},z,{\bf r}_{||}^{\prime},d\right),\;  0\leq z \leq d \;\;\;\;,\\
   G_{I}^{\left(0\right)}\left({\bf q},z,{\bf r}_{||}^{\prime},d\right),\; z\leq 0.
	\label{G0-solution}
  \end{cases}
\end{equation}
The functions $G_{I}^{\left(0\right)}$,
$G_{II}^{\left(0\right)}$ and
$G_{III}^{\left(0\right)}$ are explicitly exhibited in Appendix \ref{formulas}. 

To solve Eq. (\ref{G1-equation-fourier}), we take into account a set of four equations describing the boundary conditions for 
$G^{(1)}$. A first pair of equations is given by (see Appendix \ref{app-bc}): 
\begin{equation}
G^{(1)}\left({\bf q},d^{+},{\bf r}_{||}^{\prime},d\right)=G^{(1)}\left({\bf q},d^{-},{\bf r}_{||}^{\prime},d\right),
\label{bc-G1-1}
\end{equation}
\begin{multline}
G^{(1)}\left({\bf q},0^{+},{\bf r}^{\prime}\right)-G^{(1)}\left({\bf q},0^{-},{\bf r}^{\prime}\right)
\\
=
\int\frac{1}{\left(2\pi\right)^{2}}d^{2}{\bf q^{\prime}}h\left({\bf q}-{\bf q^{\prime}}\right)
\\
\left[-\left(\frac{\partial G^{\left(0\right)}}{\partial z}\right)\left({\bf q}^{\prime},0^{+},{\bf r}^{\prime}\right)+\left(\frac{\partial G^{\left(0\right)}}{\partial z}\right)\left({\bf q}^{\prime},0^{-},{\bf r}^{\prime}\right)\right].
\label{bc-G1-2}
\end{multline}
Then, the initial problem of finding $G$
via Eq. (\ref{gauss-law-fourier}) with the  boundary conditions (\ref{bc-main-1}) 
and (\ref{bc-main-2}) (both requiring the continuity of $G$,
with the former taken on a planar and the latter taken on a nonplanar surface),
is now effectively replaced (up to first order in $\lambda$)
by the problem of finding $G^{\left(1\right)}$ via Eq. (\ref{G1-equation-fourier})
with the boundary conditions (\ref{bc-G1-1}) and (\ref{bc-G1-2}), 
which are both taken on planar surfaces, but with the latter 
showing a discontinuity of $G^{\left(1\right)}$ when it passes through $z=0$.

Looking for a boundary condition for the $z$-derivative of $G^{\left(1\right)}$ across $z=0$, 
we integrate Eq. (\ref{G1-equation-fourier}) in $z$ in the regions $(-\eta, +\eta)$, sending $\eta\rightarrow 0$, obtaining
\begin{multline}
-\epsilon_{2}\left[\frac{\partial}{\partial z}G^{(1)}\left({\bf q},z,{\bf r}^{\prime}\right)\right]_{z=0^{+}}+\epsilon_{1}\left[\frac{\partial}{\partial z}G^{(1)}\left({\bf q},z,{\bf r}^{\prime}\right)\right]_{z=0^{-}} 
\\
=
\epsilon_{21}^{-}\int\frac{1}{\left(2\pi\right)^{2}}d^{2}{\bf q^{\prime}}h\left({\bf q}-{\bf q^{\prime}}\right)\left({\bf q^{\prime}}\cdot{\bf q}\right)G^{(0)}\left({\bf q^{\prime}},0,{\bf r^{\prime}}\right).
\label{bc-G1-3}
\end{multline}
Repeating the procedure for the region $(d-\eta, d+\eta)$, we get
\begin{multline}
\epsilon_{3}\frac{\partial}{\partial z}G^{(1)}\left({\bf q},z,{\bf r}^{\prime}\right)_{d+\eta}-\epsilon_{2}\frac{\partial}{\partial z}G^{(1)}\left({\bf q},z,{\bf r}^{\prime}\right)_{d-\eta}=0. 
\label{bc-G1-4}
\end{multline}

Requiring that $\lim_{z\rightarrow \pm \infty}G^{(1)}=0$,  we write the solution for (\ref{G1-equation-fourier}),
(\ref{bc-G1-1}), (\ref{bc-G1-2}), (\ref{bc-G1-3}) and (\ref{bc-G1-4}) as:
\begin{equation}
 G^{\left(1\right)}\left({\bf q},z,{\bf r}^{\prime}\right) = 
  \begin{cases} 
   G_{III}^{\left(1\right)}\left({\bf q},z,{\bf r}_{||}^{\prime},d\right),\; z \geq d \\
   G_{II}^{\left(1\right)}\left({\bf q},z,{\bf r}_{||}^{\prime},d\right),\;  0 < z\leq d\;\;, \\
   G_{I}^{\left(1\right)}\left({\bf q},z,{\bf r}_{||}^{\prime},d\right),\; z<0
	\label{G1-solution}
  \end{cases}
\end{equation}
with $G_{I}^{\left(1\right)}$, $G_{II}^{\left(1\right)}$ and
$G_{III}^{\left(1\right)}$ shown in Appendix \ref{formulas}. 
The solution for $G^{\left(1\right)}$ in terms of  $x$ and $y$ 
is given by (see Appendix \ref{G-1-x-y}):
\begin{multline}
G_{III}^{\left(1\right)}\left({\bf r}_{||},z,{\bf r}_{||}^{\prime},d\right)
=
\frac{1}{4\pi}\epsilon_{21}^{-}	\int d^{2}{\bf \tilde{r}}h\left(\tilde{{\bf r}}_{||}\right)
\\
\bigg[\mathcal{G}_{1}\left({\bf r}_{||},d-z,{\bf r}_{||}^{\prime},d,{\bf \tilde{r}}_{||}\right)
\\
+\frac{\epsilon_{1}}{\epsilon_{2}}\mathcal{G}_{2}\left({\bf r}_{||},d-z,{\bf r}_{||}^{\prime},d,{\bf \tilde{r}}_{||}\right)\bigg]
,\;\;\;\;\;\;\;\;\;\;
\label{G_{III}-general}
\end{multline}
\begin{multline}
G_{II}^{\left(1\right)}\left({\bf r}_{||},z,{\bf r}_{||}^{\prime},d\right)
=
\frac{1}{8\pi}\frac{\epsilon_{21}^{-}}{\epsilon_{2}}\int d^{2}{\bf \tilde{r}}h\left(\tilde{{\bf r}}_{||}\right)
\Bigg\{
\\
\epsilon_{23}^{-}\bigg[
\mathcal{G}_{1}\left({\bf r}_{||},z-d,{\bf r}_{||}^{\prime},d,{\bf \tilde{r}}_{||}\right)+\frac{\epsilon_{1}}{\epsilon_{2}}\mathcal{G}_{2}\left({\bf r}_{||},z-d,{\bf r}_{||}^{\prime},d,{\bf \tilde{r}}_{||}\right)\bigg]
\\
+\epsilon_{23}^{+}\bigg[
\mathcal{G}_{1}\left({\bf r}_{||},d-z,{\bf r}_{||}^{\prime},d,{\bf \tilde{r}}_{||}\right)
\\
+\frac{\epsilon_{1}}{\epsilon_{2}}\mathcal{G}_{2}\left({\bf r}_{||},d-z,{\bf r}_{||}^{\prime},d,{\bf \tilde{r}}_{||}\right)\bigg]\Bigg\}
,\;\;\;\;\;\;\;\;\;\;
\label{G_{II}-general}
\end{multline}
\begin{multline}
G_{I}^{\left(1\right)}\left({\bf r}_{||},z,{\bf r}_{||}^{\prime},d\right)
=
\frac{1}{8\pi}\frac{\epsilon_{21}^{-}}{\epsilon_{2}}\int d^{2}{\bf \tilde{r}}h\left(\tilde{{\bf r}}_{||}\right)
\\
\Bigg\{ \epsilon_{23}^{-}\bigg[ \mathcal{G}_{1}\left({\bf r}_{||},z-d,{\bf r}_{||}^{\prime},d,{\bf \tilde{r}}_{||}\right)+\mathcal{G}_{2}\left({\bf r}_{||},z-d,{\bf r}_{||}^{\prime},d,{\bf \tilde{r}}_{||}\right)\bigg] 
\\
+\epsilon_{23}^{+}\bigg[ \mathcal{G}_{1}\left({\bf r}_{||},z-d,{\bf r}_{||}^{\prime},d,{\bf \tilde{r}}_{||}\right)
\\
-\mathcal{G}_{2}\left({\bf r}_{||},z-d,{\bf r}_{||}^{\prime},d,{\bf \tilde{r}}_{||}\right)\bigg]\Bigg\}
,\;\;\;\;\;\;\;\;\;\;
\label{G_{I}-general}
\end{multline}
where 
\begin{multline}
\mathcal{G}_{1}\left({\bf r}_{||},\zeta,{\bf r}_{||}^{\prime},d,{\bf \tilde{r}}_{||}\right)
\\
={\bf \boldsymbol{\nabla}_{||}}G_{I}^{\left(0\right)}\left({\bf \tilde{r}}_{||},\zeta,{\bf r}_{||},d\right)\cdot{\bf \boldsymbol{\nabla}_{||}^{\prime}}G_{I}^{\left(0\right)}\left(\tilde{{\bf r}}_{||},0,{\bf r}_{||}^{\prime},d\right),
\label{G1}
\end{multline}
\begin{multline}
\mathcal{G}_{2}\left({\bf r}_{||},\zeta,{\bf r}_{||}^{\prime},d,{\bf \tilde{r}}_{||}\right)
=\left[\frac{\partial}{\partial\tilde{z}}G_{I}^{\left(0\right)}\left({\bf \tilde{r}}_{||},\tilde{z},{\bf r}_{||},d\right)\right]_{\tilde{z}=\zeta}
\\
\;\;\;\;\;\;\left[\frac{\partial}{\partial\tilde{z}}G_{I}^{\left(0\right)}\left(\tilde{{\bf r}}_{||},\tilde{z},{\bf r}_{||}^{\prime},d\right)\right]_{\tilde{z}=0}.
\label{G2}
\end{multline}
We have, therefore, concluded the solution of the problem in its most general form.
%%%%%%%%%%%%%%%%%%%%%%%%%%%%%%%%%%%%%%%%%%%%%%%%%%%%%%%%%%%%%%%
\subsection{Particular results} 

If we consider the case vacuum-dielectric ($\epsilon_3=\epsilon_2=1$), we get 
$G^{\left(1\right)}_{III}\left({\bf r}_{||},z,{\bf r}_{||}^{\prime},d\right)=G^{\left(1\right)}_{II}\left({\bf r}_{||},z,{\bf r}_{||}^{\prime},d\right)$, so that
\begin{multline}
G^{\left(1\right)}_{II}\left({\bf r}_{||},z,{\bf r}_{||}^{\prime},d\right)=\frac{1}{4\pi}\left(1-\epsilon_{1}\right)\int d^{2}{\bf \tilde{r}}h\left(\tilde{{\bf r}}_{||}\right)
\\
\Bigg\{{\bf \boldsymbol{\nabla}_{||}}G_{I}^{\left(0\right)}\left({\bf \tilde{{\bf r}}_{||}},d-z,{\bf r}_{||},d\right)\cdot{\bf \boldsymbol{\nabla}_{||}^{\prime}}G_{I}^{\left(0\right)}\left(\tilde{{\bf r}}_{||},0,{\bf r}_{||}^{\prime},d\right)
\\
+\epsilon_{1}\left[\frac{\partial}{\partial\tilde{z}}G_{I}^{\left(0\right)}\left({\bf \tilde{{\bf r}}_{||}},\tilde{z},{\bf r}_{||},d\right)\right]_{\tilde{z}=d-z}
\\
\left[\frac{\partial}{\partial\tilde{z}}G_{I}^{\left(0\right)}\left(\tilde{{\bf r}}_{||},\tilde{z},{\bf r}_{||}^{\prime},d\right)\right]_{\tilde{z}=0}
\Bigg\},\;\;\;\;\;\;\;\;\;\;
\label{G_{II}-vacuum-dielectric}
\end{multline}
\begin{multline}
G^{\left(1\right)}_{I}\left({\bf r}_{||},z,{\bf r}_{||}^{\prime},d\right)=\frac{1}{4\pi}\left(1-\epsilon_{1}\right)\int d^{2}{\bf \tilde{r}}h\left(\tilde{{\bf r}}_{||}\right)
\\
\Bigg\{{\bf \boldsymbol{\nabla}_{||}}G_{I}^{\left(0\right)}\left({\bf \tilde{r}}_{||},z+d,{\bf r}_{||},d\right)\cdot{\bf \boldsymbol{\nabla}_{||}^{\prime}}G_{I}^{\left(0\right)}\left(\tilde{{\bf r}}_{||},0,{\bf r}_{||}^{\prime},z^{\prime}\right)
\\
-\left[\frac{\partial}{\partial\tilde{z}}G_{I}^{\left(0\right)}\left({\bf \tilde{r}}_{||},\tilde{z},{\bf r}_{||},z^{\prime}\right)\right]_{\tilde{z}=z+d}
\\
\left[\frac{\partial}{\partial\tilde{z}}G_{I}^{\left(0\right)}\left(\tilde{{\bf r}}_{||},\tilde{z},{\bf r}_{||}^{\prime},d\right)\right]_{\tilde{z}=0}
\Bigg\},\;\;\;\;\;\;\;\;\;\;
\label{G_{I}-vacuum-dielectric}
\end{multline}
where, for this case,
\begin{multline}
G_{I}^{\left(0\right)}\left({\bf r}_{||},z,{\bf r}_{||}^{\prime},z^{\prime}\right)=\frac{2}{\left(\epsilon_{1}+1\right)}\frac{1}{\left[\vert{\bf r}_{||}-{\bf r}_{||}^{\prime}\vert^{2}+\left(z-d\right)^{2}\right]^{1/2}}. 
\end{multline}
The particular result given by Eqs. (\ref{G_{II}-vacuum-dielectric}) and (\ref{G_{I}-vacuum-dielectric})
also generalize that found in the literature \cite{Clinton-1985-II}. Finally, if we consider
the vacuum-conducting case ($\epsilon_3=\epsilon_2=1$ and $\epsilon_1\rightarrow \infty$),
we get
\begin{multline}
G^{\left(1\right)}_{II}\left({\bf r}_{||},z,{\bf r}_{||}^{\prime},d\right)=-\frac{1}{4\pi}\int d^{2}{\bf \tilde{r}}h\left(\tilde{{\bf r}}_{||}\right)
\\
\;\;\;\;\;\;\;\;\;\;
\left[\frac{\partial}{\partial\tilde{z}}G_{I}^{\left(0\right)}\left({\bf \tilde{{\bf r}}_{||}},\tilde{z},{\bf r}_{||},d\right)\right]_{\tilde{z}=d-z}
\\
\left[\frac{\partial}{\partial\tilde{z}}G_{I}^{\left(0\right)}\left(\tilde{{\bf r}}_{||},\tilde{z},{\bf r}_{||}^{\prime},d\right)\right]_{\tilde{z}=0},\;\;\;\;\;\;\;\;
\label{G_{II}-vacuum-conductor}
\end{multline}
\begin{equation}
G^{\left(1\right)}_{I}\left({\bf r}_{||},z,{\bf r}_{||}^{\prime},d\right)=0,
\label{G_{I}-vacuum-conductor}
\end{equation}
where, for this case,
\begin{multline}
G_{I}^{\left(0\right)}\left({\bf r}_{||},z,{\bf r}_{||}^{\prime},d\right)=\frac{2}{\left[\vert{\bf r}_{||}-{\bf r}_{||}^{\prime}\vert^{2}+\left(z-d\right)^{2}\right]^{1/2}}. 
\end{multline}
This result recovers the result found in the literature \cite{Clinton-1985-II}
and is formally identical to Hadamard's theorem for Green's functions \cite{Clinton-1985-II, Hadamard-1910}, which gives 
the solution (up to first order in $\lambda$) of
\begin{equation}
\boldsymbol{\nabla}^2{G\left({\bf r},{\bf r}^{\prime}\right)}=-4\pi \delta\left({\bf r}-{\bf r}^{\prime}\right),
\label{hadamard-1}
\end{equation}
with the boundary condition
\begin{equation}
G\left({\bf r}-{\bf r}^{\prime}\right)\vert_{z=\lambda h({\bf r}_{||})}=0.
\label{hadamard-2}
\end{equation}

%%%%%%%%%%%%%%%%%%%%%%%%%%%%%%%%%%%%%%%%%%%%%%%%%%%%%%%%%%%%%%%
\subsection{Interaction between a charge and the surrounding polarized matter induced by it} 

When we bring a charge $Q$ to the position ${\bf r}^{\prime}$, we produce a state of polarization 
in the dielectric media. The energy of interaction $W$ between the charge $Q$ and the 
polarized  dielectrics (we are considering a linear behavior for the dielectrics)
is given by (see Appendix \ref{app-W}):
\begin{equation}
W=\frac{1}{2}Q\phi_{ind}\left({\bf r}^{\prime}\right),
\label{W-phi-ind}
\end{equation}
where $\phi_{ind}$ is the induced (or image) potential function, which, taken
at ${\bf r}^{\prime}$, is given by
\begin{equation}
\phi_{ind}\left({\bf r}^{\prime}\right)\approx Q\left[G_{ind}^{(0)}\left({\bf r},{\bf r^{\prime}}\right)_{{\bf r}={\bf r}^{\prime}}+\lambda G^{(1)}\left({\bf r},{\bf r^{\prime}}\right)_{{\bf r}={\bf r}^{\prime}}\right].
\end{equation}
Then, we have
\begin{equation}
W\approx W^{(0)}+\lambda W^{(1)},
\label{W-W0-W1}
\end{equation}
where
\begin{eqnarray}
W^{(0)}&=&\frac{1}{2}Q^{2}G_{ind}^{(0)}\left({\bf r},{\bf r^{\prime}}\right)_{{\bf r}={\bf r}^{\prime}},
\cr\cr
W^{(1)}&=&\frac{1}{2}Q^{2}G^{(1)}\left({\bf r},{\bf r^{\prime}}\right)_{{\bf r}={\bf r}^{\prime}},
\label{W-G0-G1}
\end{eqnarray}
with
\begin{eqnarray}
G_{ind}^{\left(0\right)}\left({\bf r}_{||},z,{\bf r}_{||}^{\prime},d\right)
&=&
-\int\frac{1}{\left(2\pi\right)^{2}}d^{2}{\bf q}e^{-qz}\frac{1}{q}e^{i{\bf q}\cdot{\bf r}_{||}}e^{-i{\bf q}\cdot{\bf r}_{||}^{\prime}}
\nonumber
\\
&&\frac{1}{\epsilon_{23}^{+}}\frac{8\pi\epsilon_{2}\epsilon_{12}^{-}}{\left[\epsilon_{23}^{-}\epsilon_{12}^{-}e^{-qd}
+\epsilon_{23}^{+}\epsilon_{12}^{+}e^{qd}\right]}.
\end{eqnarray}
Considering the solution for $G^{(1)}$ for $z=d$, we have:
\begin{multline}
W=-Q^{2}\frac{1}{\pi}\frac{\epsilon_{2}\epsilon_{12}^{-}}{\epsilon_{23}^{+}}
\int d^{2}{\bf q}\frac{1}{q}\frac{e^{-qd}}{\left[\epsilon_{23}^{-}\epsilon_{12}^{-}e^{-qd}+\epsilon_{23}^{+}\epsilon_{12}^{+}e^{qd}\right]}
\\
+\lambda Q^{2}\epsilon_{21}^{-}\frac{1}{8\pi}\int d^{2}{\bf \tilde{r}}h\left(\tilde{{\bf r}}_{||}\right)
\bigg\{ {\bf \vert\boldsymbol{\nabla}_{||}^{\prime}}G_{I}^{\left(0\right)}\left(\tilde{{\bf r}}_{||},0,{\bf r}_{||}^{\prime},d\right)\vert^{2}
\\
+\frac{\epsilon_{1}}{\epsilon_{2}}\left(\left[\frac{\partial}{\partial\tilde{z}}G_{I}^{\left(0\right)}\left(\tilde{{\bf r}}_{||},\tilde{z},{\bf r}_{||}^{\prime},d\right)\right]_{\tilde{z}=0}\right)^{2}\bigg\}.  
\label{W-3D-forma-1}
\end{multline}
If we consider the case vacuum-dielectric ($\epsilon_3=\epsilon_2=1$) in Eq. (\ref{W-3D-forma-1}), 
we obtain
\begin{multline}
W=-Q^{2}\frac{1}{4d}\frac{\left(\epsilon_{1}-1\right)}{\left(\epsilon_{1}+1\right)}
\\
+\lambda Q^{2}\frac{(1-\epsilon_1)}{8\pi}\int d^{2}{\bf \tilde{r}}h\left(\tilde{{\bf r}}_{||}\right)
\bigg\{ {\bf \vert\boldsymbol{\nabla}_{||}^{\prime}}G_{I}^{\left(0\right)}\left(\tilde{{\bf r}}_{||},0,{\bf r}_{||}^{\prime},d\right)\vert^{2}
\\
+{\epsilon_{1}}\left(\left[\frac{\partial}{\partial\tilde{z}}G_{I}^{\left(0\right)}\left(\tilde{{\bf r}}_{||},\tilde{z},{\bf r}_{||}^{\prime},d\right)\right]_{\tilde{z}=0}\right)^{2}\bigg\},  
\label{W-3D-forma-1-vacuum-dielectric}
\end{multline}
where $G_{I}^{\left(0\right)}$ for this case is obtained considering $\epsilon_3=\epsilon_2=1$
in Eqs. (\ref{GI0}) and (\ref{A0}), and using (\ref{fourier-representation}).
The result shown in Eq. (\ref{W-3D-forma-1-vacuum-dielectric}) is agreement with that found
in the literature \cite{Clinton-1985-II}.
For the vacuum-conducting case ($\epsilon_3=\epsilon_2=1$ and $\epsilon_1\rightarrow -\infty$), we obtain
from Eq. (\ref{W-3D-forma-1}) the result
\begin{multline}
W=-Q^{2}\frac{1}{4d}
\\
-\lambda Q^{2}\frac{1}{8\pi}\int d^{2}{\bf \tilde{r}}h\left(\tilde{{\bf r}}_{||}\right)
\bigg\{\left(\left[\frac{\partial}{\partial\tilde{z}}G_{I}^{\left(0\right)}\left(\tilde{{\bf r}}_{||},\tilde{z},{\bf r}_{||}^{\prime},d\right)\right]_{\tilde{z}=0}\right)^{2}\bigg\},  
\label{W-3D-forma-1-vacuum-conductor	}
\end{multline}

We can also rewrite Eq. (\ref{W-3D-forma-1}) in the following manner:
\begin{multline}
W=-Q^{2}\frac{\epsilon_{2}\epsilon_{12}^{-}}{\epsilon_{23}^{+}}\frac{1}{d}\mathcal{F}_{1}\left(\epsilon_{23}^{-}\epsilon_{12}^{-},\epsilon_{23}^{+}\epsilon_{12}^{+}\right)
\\
+\lambda Q^{2}\epsilon_{21}^{-}\frac{1}{\pi^{2}}\int d^{2}{\bf \tilde{q}}h\left({\bf \tilde{q}}\right)e^{i\tilde{{\bf q}}\cdot{\bf r}_{||}^{\prime}}\int_{0}^{\infty}dR\;RJ_{0}\left(\tilde{q}R\right)\int_{0}^{\infty}dq
\\
\int_{0}^{\infty}dq^{\prime}{\cal F}_{2}\left(q,\epsilon_{23}^{-}\epsilon_{12}^{-},\epsilon_{23}^{+}\epsilon_{12}^{+},d\right){\cal F}_{2}\left(q^{\prime},\epsilon_{23}^{-}\epsilon_{12}^{-},\epsilon_{23}^{+}\epsilon_{12}^{+},d\right) 
\\
\times\left\{ \epsilon_{2}^{2}J_{1}\left(qR\right)J_{1}\left(q^{\prime}R\right)+\epsilon_{1}\epsilon_{2}J_{0}\left(qR\right)J_{0}\left(q^{\prime}R\right)\right\},
\label{W-form-2}
\end{multline} 
where
\begin{equation}
\mathcal{F}_{1}\left(a,b\right)=
\begin{cases}
\begin{array}{cc}
[\ln\left(a+b\right)-\ln\left(b\right)]/a & a\neq0\\
{1}/{b} & a=0,
\end{array}
\end{cases}
\end{equation}
\begin{equation}
{\cal F}_{2}\left(q,a,b,d\right)={q}/{\left(ae^{-qd}+be^{qd}\right)},
\end{equation}
and $J_0$ and $J_1$ are Bessel functions of the first kind
of zeroth and first order, respectively.
The form for $W$ given in Eq. (\ref{W-form-2}) is very convenient for numerical
calculations.
Next, we take some limits of the presented formulas and recover some results
found in the literature.

For the case with
$\epsilon_{3}=\epsilon_{2}$, we have the problem involving
two dielectrics $\epsilon_2$ and $\epsilon_1$, for which we get:
\begin{multline}
W=Q^{2}\frac{\epsilon_{21}^{-}}{\epsilon_{12}^{+}}\frac{1}{4d}\frac{1}{\epsilon_{2}}
+\lambda Q^{2}\frac{\epsilon_{21}^{-}}{\epsilon_{2}{(\epsilon_{12}^{+})}^{2}}\frac{1}{4\pi^{2}}\int d^{2}{\bf \tilde{q}}h\left({\bf \tilde{q}}\right)e^{i\tilde{{\bf q}}\cdot{\bf r}_{||}^{\prime}}
\\
\int_{0}^{\infty}dRRJ_{0}\left(\tilde{q}R\right)\left\{ \frac{\epsilon_{2}R^{2}+\epsilon_{1}d^{2}}{\left(R^{2}+d^{2}\right)^{3}}\right\} 
\label{W-2D}
\end{multline}
The vacuum-dielectric case can be recovered by doing
$\epsilon_2=1$ in Eq. (\ref{W-2D}). The vacuum-conducting case is recovered taking $\epsilon_1 \rightarrow -\infty$.
Both results for these limit cases coincide with those found in the literature \cite{Clinton-1985-I,Clinton-1985-II}.

The presence of a nonplanar
interface between $\epsilon_2$ and $\epsilon_1$ media also induces an external field, so that on each
charge $Q$ in the two-dimensional system acts an effective force parallel to the plane $z=d$ given by
\begin{equation}
{\bf F}_{||}^{(ext)}\approx-\frac{1}{2}\lambda Q^{2}\boldsymbol{\nabla}_{||}^{\prime}G^{(1)}\left({\bf r}_{||}^{\prime},d,{\bf r}_{||}^{\prime},d\right).
\label{F-||-ext}
\end{equation}
This force depends on the magnitude of the charge (specifically, on $Q^2$) and
it can point to the next valley or peak of the nonplanar interface, 
depending on the sign of $\epsilon_2-\epsilon_1$.
This generalizes the result found in the literature for the case vacuum-conductor \cite{Clinton-1985-I},
where the correspondent force always points to the next peak of
the nonplanar interface.

We also have a perpendicular force acting on $Q$, given by
\begin{eqnarray}
{\bf F}_{\perp}^{(ext)}&\approx&-\frac{1}{2}Q^{2}\frac{\partial}{\partial d}
\Bigg[G_{ind}^{(0)}\left({\bf r}_{||}^{\prime},d,{\bf r}_{||}^{\prime},d\right)
\cr\cr
&&+\lambda G^{(1)}\left({\bf r}_{||}^{\prime},d,{\bf r}_{||}^{\prime},d\right)\Bigg].
\label{F-perpend-ext}
\end{eqnarray}
Part of this force (proportional to $\lambda$) can be quite relevant for suspended graphene, since the force induces a deformation of the material which, in turn, affects its optical and DC transport properties.

%%%%%%%%%%%%%%%%%%%%%%%%%%%

%%%%%%%%%%%%%%%%%%%%%%%%%%%

%%%%%%%%%%%%%%%%%%%%%%%%%%%%%%%%%%%%%%%%%%%%%%%%%%%%%%%%%%%%%%%
\subsection{Effective charge-charge interaction in the 2D-material} 

Let us now consider the effective electron-electron interaction, which alters its usual form due to the presence of corrugation. 
The effective electric potential $\phi^{(eff)}$ associated to a point-charge
$Q$ in the position ${\bf r}_{||}^{\prime}$ is 
\begin{equation}
\phi^{(eff)}\left({\bf r}_{||},{\bf r}_{||}^{\prime},d\right)=\phi\left({\bf r}_{||},d,{\bf r}_{||}^{\prime},d\right)=QG\left({\bf r}_{||},d,{\bf r}_{||}^{\prime},d\right).
\label{phi-eff-exato}
\end{equation}
Up to the first order, we have
\begin{equation}
\phi^{(eff)}\left({\bf r}_{||},{\bf r}_{||}^{\prime},d\right)\approx\phi^{\left(0\right)}\left({\bf r}_{||},d,{\bf r}_{||}^{\prime},d\right)+\lambda\phi^{\left(1\right)}\left({\bf r}_{||},d,{\bf r}_{||}^{\prime},d\right).
\label{phi-eff-0-1}
\end{equation}
Using Eqs. (\ref{G0-solution}) and (\ref{G1-solution}), we get
\begin{multline}
\phi^{(eff)}\left({\bf q},{\bf r}_{||}^{\prime},d\right)\approx\frac{2\pi Q}{q}e^{-i{\bf q}\cdot{\bf r}_{||}^{\prime}}\bigg\{
%\\
\frac{2\left[\epsilon_{21}^{-}e^{-2qd}+\epsilon_{12}^{+}\right]}{\left[\epsilon_{23}^{-}\epsilon_{12}^{-}e^{-2qd}+\epsilon_{23}^{+}\epsilon_{12}^{+}\right]}
\\
+\lambda\frac{\epsilon_{2}\left(\tilde{T}_{2}\epsilon_{21}^{-}+\tilde{T}_{1}\epsilon_{1}q\right)e^{qd}}{q\left[\epsilon_{23}^{-}\epsilon_{12}^{-}e^{-qd}+\epsilon_{23}^{+}\epsilon_{12}^{+}e^{qd}\right]}\bigg\},
\label{phi-eff-fourier}
\end{multline}
where
\begin{eqnarray}
\tilde{T}_{1}&=&\int\frac{1}{\left(2\pi\right)^{2}}d^{2}{\bf q^{\prime}}
%\\
\frac{4qh\left({\bf q}-{\bf q^{\prime}}\right)\epsilon_{21}^{-}e^{i{\bf r}_{||}^{\prime}\cdot\left({\bf q}-{\bf q}^{\prime}\right)}}{\left[\epsilon_{23}^{-}\epsilon_{12}^{-}e^{-q^{\prime}d}+\epsilon_{23}^{+}\epsilon_{12}^{+}e^{q^{\prime}d}\right]},
\\
\nonumber
\\
\tilde{T}_{2}&=&\int\frac{1}{\left(2\pi\right)^{2}}d^{2}{\bf q^{\prime}}
\frac{4qh\left({\bf q}-{\bf q^{\prime}}\right)\left({\bf q^{\prime}}\cdot{\bf q}\right)\epsilon_{2}e^{i{\bf r}_{||}^{\prime}\cdot\left({\bf q}-{\bf q}^{\prime}\right)}}{q^{\prime}\left[\epsilon_{23}^{-}\epsilon_{12}^{-}e^{-q^{\prime}d}+\epsilon_{23}^{+}\epsilon_{12}^{+}e^{q^{\prime}d}\right]}.
\end{eqnarray}
From Eq. (\ref{phi-eff-fourier}), we get the effective dielectric constant $\epsilon_{eff}$:
\begin{eqnarray}
\frac{1}{\epsilon_{eff}}&=&\frac{2\left[\epsilon_{21}^{-}e^{-2qd}+\epsilon_{12}^{+}\right]}{\left[\epsilon_{23}^{-}\epsilon_{12}^{-}e^{-2qd}+\epsilon_{23}^{+}\epsilon_{12}^{+}\right]}
\nonumber
\\
&&+\lambda\frac{\epsilon_{2}\left(\tilde{T}_{2}\epsilon_{21}^{-}+\tilde{T}_{1}\epsilon_{1}q\right)e^{qd}}{q\left[\epsilon_{23}^{-}\epsilon_{12}^{-}e^{-qd}+\epsilon_{23}^{+}\epsilon_{12}^{+}e^{qd}\right]}.
\label{epsilon-eff}
\end{eqnarray}
Notice that ${\epsilon_{eff}}$ depends on ${\bf r}_{||}^{\prime}$. When $\lambda=0$, the result given in Eq. (\ref{epsilon-eff}) recovers that found in the literature \cite{Profumo-2010,Katsnelson-2011,Nuno-Livro}. 
The functions $\phi^{(0)}$ and $\phi^{(1)}$ in Eq. (\ref{phi-eff-0-1}) are given
explicitly in Appendix \ref{phi-E-coordinate} and exhibit the symmetry properties 
\begin{eqnarray}
\phi^{\left(0\right)}\left({\bf r}_{||},d,{\bf r}_{||}^{\prime},d\right)
&=&\phi^{\left(0\right)}\left({\bf r}_{||}^{\prime},d,{\bf r}_{||},d\right)
\\
\phi^{\left(1\right)}\left({\bf r}_{||},d,{\bf r}_{||}^{\prime},d\right)
&=&\phi^{\left(1\right)}\left({\bf r}_{||}^{\prime},d,{\bf r}_{||},d\right),
\end{eqnarray}
from which the energy interaction $W_{12} $ between two charges $Q_1$ and  $Q_2$ located at ${\bf r}_{1||}$ and ${\bf r}_{2||}$, respectively, is
\begin{eqnarray}
W_{12}&\approx& Q_2\phi^{(eff)}_{1}\left({\bf r}_{2||},{\bf r}_{1||},d\right)
=W_{21}
\cr\cr
&\approx& Q_1\phi^{(eff)}_{2}\left({\bf r}_{1||},{\bf r}_{2||},d\right),
\end{eqnarray}
as expected, with $\phi^{(eff)}_{1}$ and $\phi^{(eff)}_{2}$ the potential functions associated to the charges
1 e 2.
	
The effective electric field ${\bf E}_{||}^{(eff)}$ produced by a charge $Q$ is 
\begin{eqnarray}
{\bf E}_{||}^{(eff)}\left({\bf r}_{||},{\bf r}_{||}^{\prime},d\right)=
-\boldsymbol{\nabla}_{||}\phi^{(eff)}\left({\bf r}_{||},{\bf r}_{||}^{\prime},d\right),
\end{eqnarray}
which can be written as
\begin{equation}
{\bf E}_{||}^{(eff)}\left({\bf r}_{||},{\bf r}_{||}^{\prime},d\right)
\approx
{\bf E}_{||}^{\left(0\right)}\left({\bf r}_{||},d,{\bf r}_{||}^{\prime},d\right)
+\lambda{\bf E}_{||}^{\left(1\right)}\left({\bf r}_{||},d,{\bf r}_{||}^{\prime},d\right),
\label{E-eff-0-1}
\end{equation}
with 
\begin{eqnarray}
{\bf E}_{||}^{\left(0\right)}\left({\bf r}_{||},{\bf r}_{||}^{\prime},d\right)&=&-\boldsymbol{\nabla}_{||}\phi^{\left(0\right)}
\left({\bf r}_{||},d,{\bf r}_{||}^{\prime},d\right),
\\
{\bf E}_{||}^{\left(1\right)}\left({\bf r}_{||},{\bf r}_{||}^{\prime},d\right)&=&-\boldsymbol{\nabla}_{||}\phi^{\left(1\right)}
\left({\bf r}_{||},d,{\bf r}_{||}^{\prime},d\right),
\end{eqnarray}
and ${\bf E}_{||}^{\left(0\right)}$ and ${\bf E}_{||}^{\left(1\right)}$ given explicitly 
in Appendix \ref{phi-E-coordinate}.

For ${\bf E}_{||}^{\left(0\right)}$, we have the usual symmetry
\begin{equation}
{\bf E}_{||}^{\left(0\right)}\left({\bf r}_{||},d,{\bf r}_{||}^{\prime},d\right)=
-{\bf E}_{||}^{\left(0\right)}\left({\bf r}_{||}^{\prime},d,{\bf r}_{||},d\right),
\label{E-0-symmetry}
\end{equation}
but for ${\bf E}_{||}^{\left(1\right)}$, in general, we find
\begin{equation}
{\bf E}_{||}^{\left(1\right)}\left({\bf r}_{||},d,{\bf r}_{||}^{\prime},d\right)\neq-{\bf E}_{||}^{\left(1\right)}\left({\bf r}_{||}^{\prime},
d,{\bf r}_{||},d\right).
\label{E-1-symmetry}
\end{equation} 
Then, the effective 2D electrical field is such that
\begin{equation}
{\bf E}_{||}^{\left(eff\right)}\left({\bf r}_{||},{\bf r}_{||}^{\prime},d\right)\neq-{\bf E}_{||}^{\left(eff\right)}\left({\bf r}_{||}^{\prime},{\bf r}_{||},d\right),
\label{E-eff-symmetry}
\end{equation}
so that the effective forces between two charges do not point along the line from one charge to the other.

\begin{figure}[t]
\begin{centering}
\includegraphics[width=0.8\columnwidth]{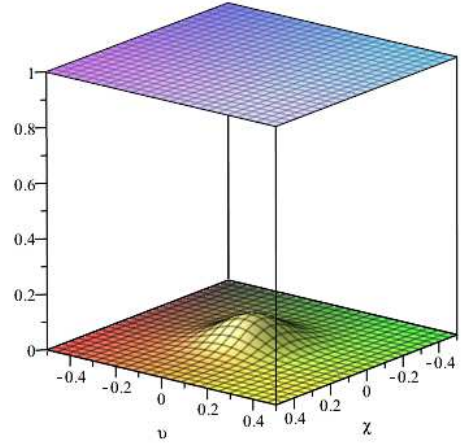} 
\end{centering}
\caption{Illustration of the planar two-dimensional system in the presence
of a nonplanar surface described by Eq. (\ref{h-gauss}), with $d=300nm$, $k=2\pi/d$,
and $\lambda=1/10$. The vertical axis exhibits $z/d$, whereas the other axes represent
$\chi=x/d$ and $\upsilon=y/d$.}
\label{case-gaussian-plane-system-nonplanar-interface} 
\end{figure}

\begin{figure}[t]
\begin{centering}
\includegraphics[width=0.6\columnwidth]{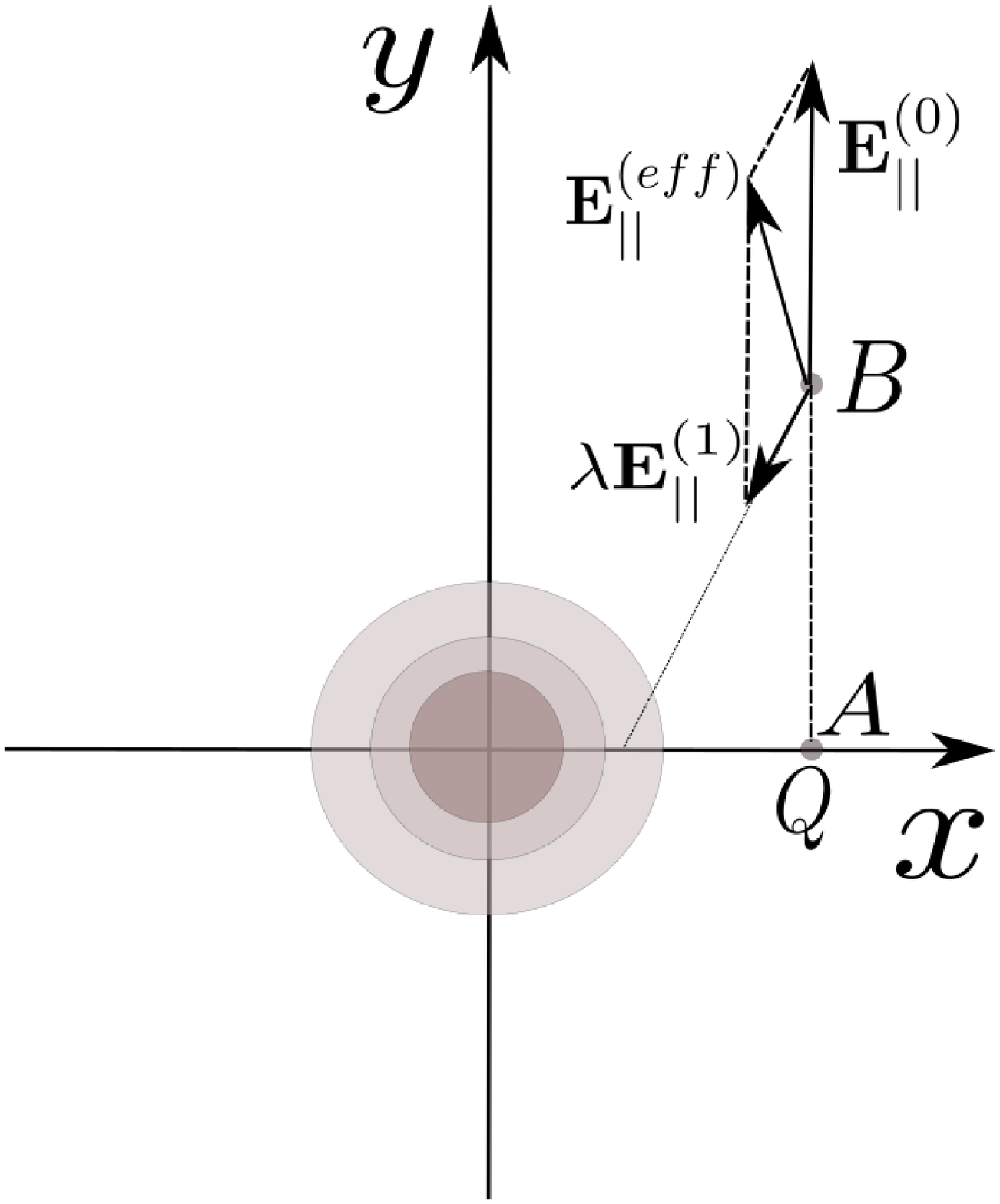} 
\end{centering}
\caption{Front view of the 2D system located on the plane $z=d$
($z$-axis is perpendicular to the paper). The circular lines are 
projections on the plane $z=d$ of some contour lines indicating points of
equal altitude of the gaussian surface shown in Fig. \ref{case-gaussian-plane-system-nonplanar-interface}.
A charge $Q>0$ is located at the point $A$, whereas
the electric field is shown at the point $B$.}
\label{curvas-nivel} 
\end{figure}

\begin{figure}[t]
\begin{centering}
\includegraphics[width=0.6\columnwidth]{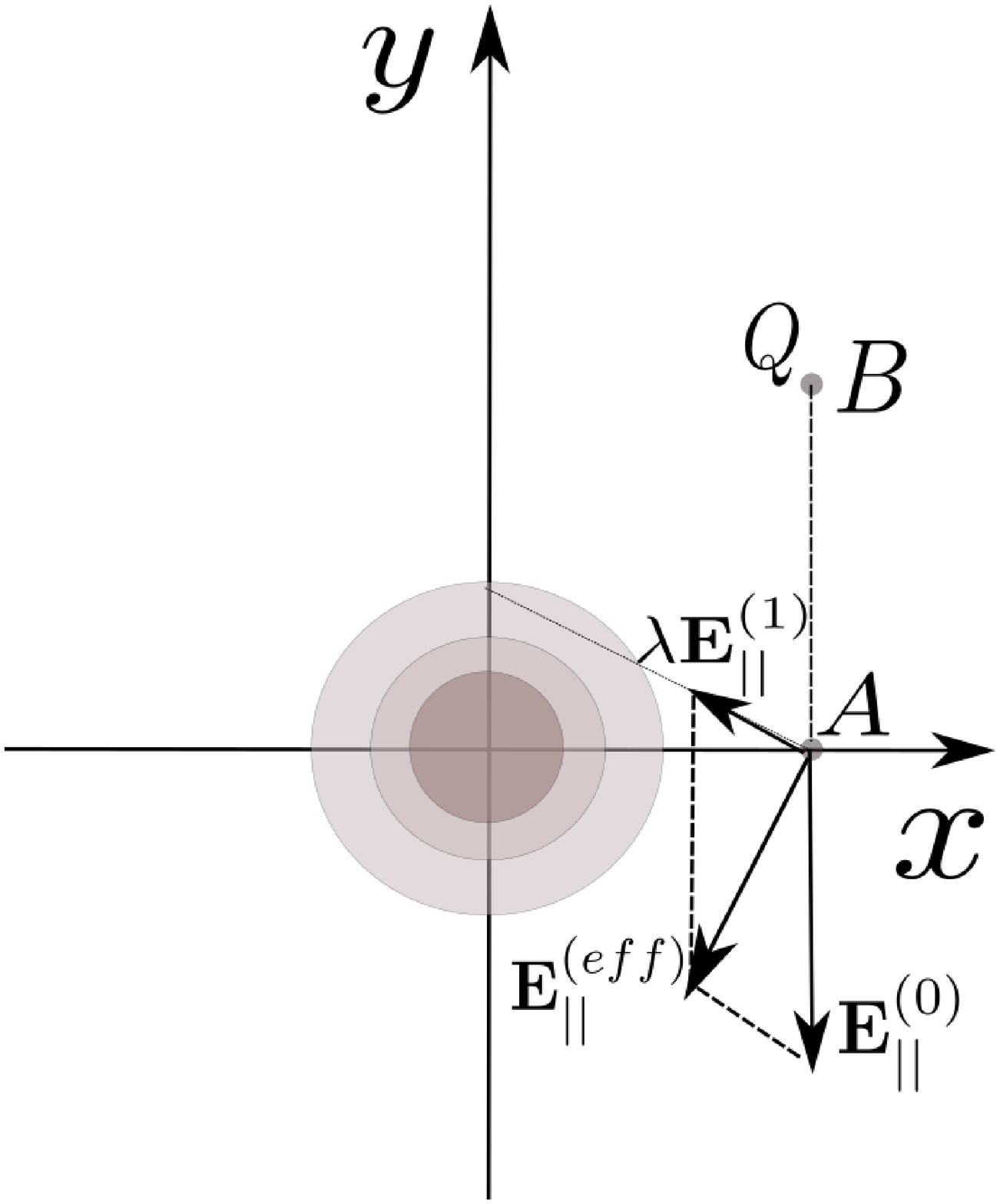} 
\end{centering}
\caption{Front view of the 2D system located on the plane $z=d$
($z$-axis is perpendicular to the paper). The circular lines are 
projections on the plane $z=d$ of some contour lines indicating points of
equal altitude of the gaussian surface shown in Fig. \ref{case-gaussian-plane-system-nonplanar-interface}.
A charge $Q>0$ is located at the point $B$, whereas
the electric field is shown at the point $A$.}
\label{curvas-nivel-reverso} 
\end{figure}

The above result can be understood as follows. 
Under the external field associated to a charge $Q$, the atoms
of the dielectric media become polarized, or with permanent dipoles 
aligned with the field \cite{Jackson-livro}.
Let us consider that these dipole moments contribute to the averaged 
charge density of the dielectric media \cite{Jackson-livro},
\begin{equation}
\langle\rho^{\prime}\left({\bf r}\right)\rangle=\frac{1}{\Delta V}\int_{\Delta V}
\rho^{\prime}\left({\bf r}+\boldsymbol{\xi}\right)d^{3}\boldsymbol{\xi},
\label{mean-charge-density}
\end{equation}
where $\rho^{\prime}$ is the exact position (in a certain instant of time)
of the charges in motion (thermal or zero point effects) in the dielectric media, 
$\Delta V$ is a macroscopically
small volume, $\boldsymbol{\xi}$ ranges over the this small volume, and $\langle\;\rangle$ means the average value \cite{Jackson-livro}.
For the situation where all interfaces are flat, $\langle\rho^{\prime}\left({\bf r}\right)\rangle$,
now relabeled as $\langle\rho^{\prime\left(0\right)}({\bf r})\rangle$, is (by symmetry arguments) 
a function of $z$ and of the distance $\vert{\bf r}_{||}-{\bf r}_{||}^{\prime}\vert$. 
Then, the effective potential $\phi^{\left(eff\right)}$, for this case,
is $\phi^{\left(eff\right)}=\phi^{\left(0\right)}$,
which is associated to the distribution of charges 
$Q\delta({\bf r}-{\bf r}^{\prime})+
\langle\rho^{\prime\left(0\right)}({\bf r})\rangle$.
From Eq. (\ref{phi-0-eff}) we can see that $\phi^{\left(0\right)}\left({\bf r}_{||},d,{\bf r}_{||}^{\prime},d\right)$ 
depends on $d$ and on the distance $\vert{\bf r}_{||}-{\bf r}_{||}^{\prime}\vert$.
This means that the equipotential  lines are circular lines with the charge $Q$ at the center of the circle. 
For this case,
the effective electric field ${\bf E}_{||}^{(eff)}$ is 
${\bf E}_{||}^{(eff)}={\bf E}_{||}^{\left(0\right)}$,
with ${\bf E}_{||}^{\left(0\right)}$ [Eq. (\ref{E-0-eff})]
proportional to the vector ${\bf r}_{||}-{\bf r}_{||}^{\prime}$,
depending on $d$, on the distance $\vert{\bf r}_{||}-{\bf r}_{||}^{\prime}\vert$,
and exhibiting the symmetry shown in Eq. (\ref{E-0-symmetry}). 
On the other hand, if we consider a nonplanar surface,
for example, the surface defined by the 2D gaussian function,
\begin{equation}
z=\lambda h\left({\bf r}_{||}\right)=\lambda de^{-k^{2}\left(x^{2}+y^{2}\right)},
\label{h-gauss}
\end{equation}
as illustrated in Fig. \ref{case-gaussian-plane-system-nonplanar-interface},
the mean charge density $\langle\rho^{\prime}\left({\bf r}\right)\rangle$
can be written as
\begin{equation}
\langle\rho^{\prime}({\bf r})\rangle
\approx\langle\rho^{\prime\left(0\right)}({\bf r})\rangle
+
\lambda\langle\rho^{\prime\left(1\right)}({\bf r})\rangle,
\label{rho-primo-0-1}
\end{equation}
where the term $\lambda\langle\rho^{\prime\left(1\right)}({\bf r})\rangle$
is related to the presence of the surface $z=\lambda h\left({\bf r}_{||}\right)$.
If a charge $Q$, in the 2D system ($z=d$), is put exactly over the center (peak) of the gaussian, 
the term $\lambda\langle\rho^{\prime\left(1\right)}({\bf r})\rangle$ in Eq. (\ref{rho-primo-0-1})
is, by the symmetry of this situation, a function of $z$ and of the distance $\vert{\bf r}_{||}-{\bf r}_{||}^{\prime}\vert$.
For this specific position of $Q$, the effective electric field 
${\bf E}_{||}^{(eff)}\approx{\bf E}_{||}^{\left(0\right)}+\lambda{\bf E}_{||}^{\left(1\right)}$
is along the line from the charge $Q$ to any other point of the plane $z=d$,
since both parts, ${\bf E}_{||}^{\left(0\right)}$ and  $\lambda{\bf E}_{||}^{\left(1\right)}$,
are proportional to ${\bf r}_{||}-{\bf r}_{||}^{\prime}$.
However, this is a particular situation. If $Q$ is not over the peak, but displaced along the $x$ axis,
as shown in Fig. \ref{curvas-nivel},
its expected that external field associated to the charge $Q$
contributes to different average
charge densities in the dielectric media for the left and right side of $Q$ , in the sense that,
in general, $\langle\rho^{\prime\left(1\right)}({\bf r}^{\prime}_{||}-\delta_x\hat{{\bf x}}+\delta_y\hat{{\bf y}}+z\hat{{\bf z}})\rangle
\neq\langle\rho^{\prime\left(1\right)}({\bf r}^{\prime}_{||}+\delta_x\hat{{\bf x}}+\delta_y\hat{{\bf y}}+z\hat{{\bf z}})\rangle$.
This means that for the potential $\phi^{(eff)}\approx\phi^{(0)}+\lambda\phi^{(1)}$,
although for the term $\phi^{(0)}$ the equipotential lines are circular lines with $Q$
at the center, those associated with $\phi^{(1)}$ are not circular lines, 
so that $\lambda{\bf E}_{||}^{\left(1\right)}$ (and, as a consequence,
${\bf E}_{||}^{(eff)}$) is not proportional to the vector ${\bf r}_{||}-{\bf r}_{||}^{\prime}$, 
and consequently is not along the line from the charge $Q$
(point $A$) to the point $B$ in Fig. \ref{curvas-nivel}.
Inversely, putting the charge $Q$ at the point $B$, by analogous arguments we expect the behavior of
${\bf E}_{||}^{(eff)}$ is as shown in Fig. \ref{curvas-nivel-reverso}.
Comparing ${\bf E}_{||}^{(eff)}$ in Fig. \ref{curvas-nivel} and Fig. \ref{curvas-nivel-reverso},
one can visualize the inequality given in Eq. (\ref{E-eff-symmetry}). Comparing the behavior
of ${\bf E}_{||}^{\left(0\right)}$ and ${\bf E}_{||}^{\left(1\right)}$, one can also
visualize Eqs. (\ref{E-0-symmetry}) and (\ref{E-1-symmetry}).

%%%%%%%%%%%%%%%%%%%%%%%%%%%%%%%%%%%%%%%%%%%%%%%%%%%%%%%%%%%%%%%%%%%%%%%%%%%%%%%%%%%%%%%%%%%%%%%%
\section{Some applications}
\label{app-point-charge}

For simplicity, let us consider the situation with $\epsilon_{3}=\epsilon_{2}$, for which can use Eq. (\ref{W-2D}), which we write
as $W=W^{(0)}+\lambda W^{(1)}$, where $W^{(0)}$ is obtained making $\lambda=0$ in Eq. (\ref{W-2D}).
For this case, and when $\lambda=0$ (the interface between $\epsilon_2$ and $\epsilon_1$ is plane), 
a perpendicular force ${\bf F}_{\bot}^{(0)}=-\frac{\partial}{\partial d}W^{(0)} \hat{{\bf z}}$
acts on a point-charge $Q$ in the two-dimensional system, which will be used as reference in
comparison to the external parallel force given in Eq. (\ref{F-||-ext}).

%%%%%%%%%%%%%%%%%%%%%%%%%%%%%%%%%%%%%%%%%%%%%%%%%%%%%%%%%%%%%%%%%%%%%%%%%%%%%%%%%%%%%%%%%%%%%%%%
\subsection{2D sine-grating}
\label{2D-sine-grating}

\begin{figure}[t]
\begin{centering}
\includegraphics[width=0.8\columnwidth]{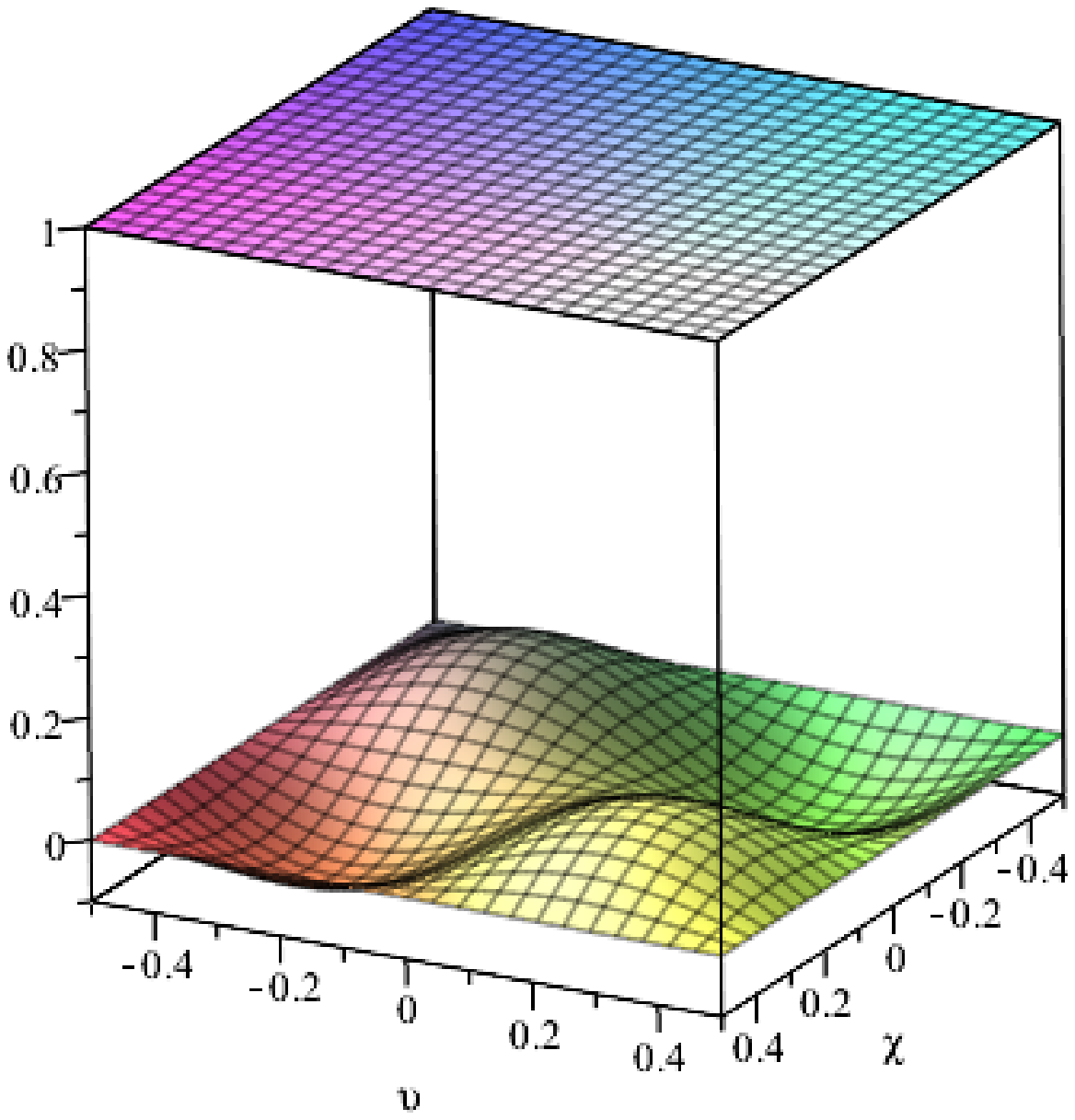} 
\end{centering}
\caption{Illustration of the planar two-dimensional system in the presence
of a nonplanar surface described by Eq. (\ref{h-sin-sin}), with $L_x=L_y=d=300nm$,
and $\lambda=1/10$. The vertical axis exhibits $z/d$, whereas the other axes represent
$\chi=x/d$ and $\upsilon=y/d$.}
\label{case-sin-sin-plane-system-nonplanar-interface} 
\end{figure}

\begin{figure}[t]
\begin{centering}
\includegraphics[width=0.8\columnwidth]{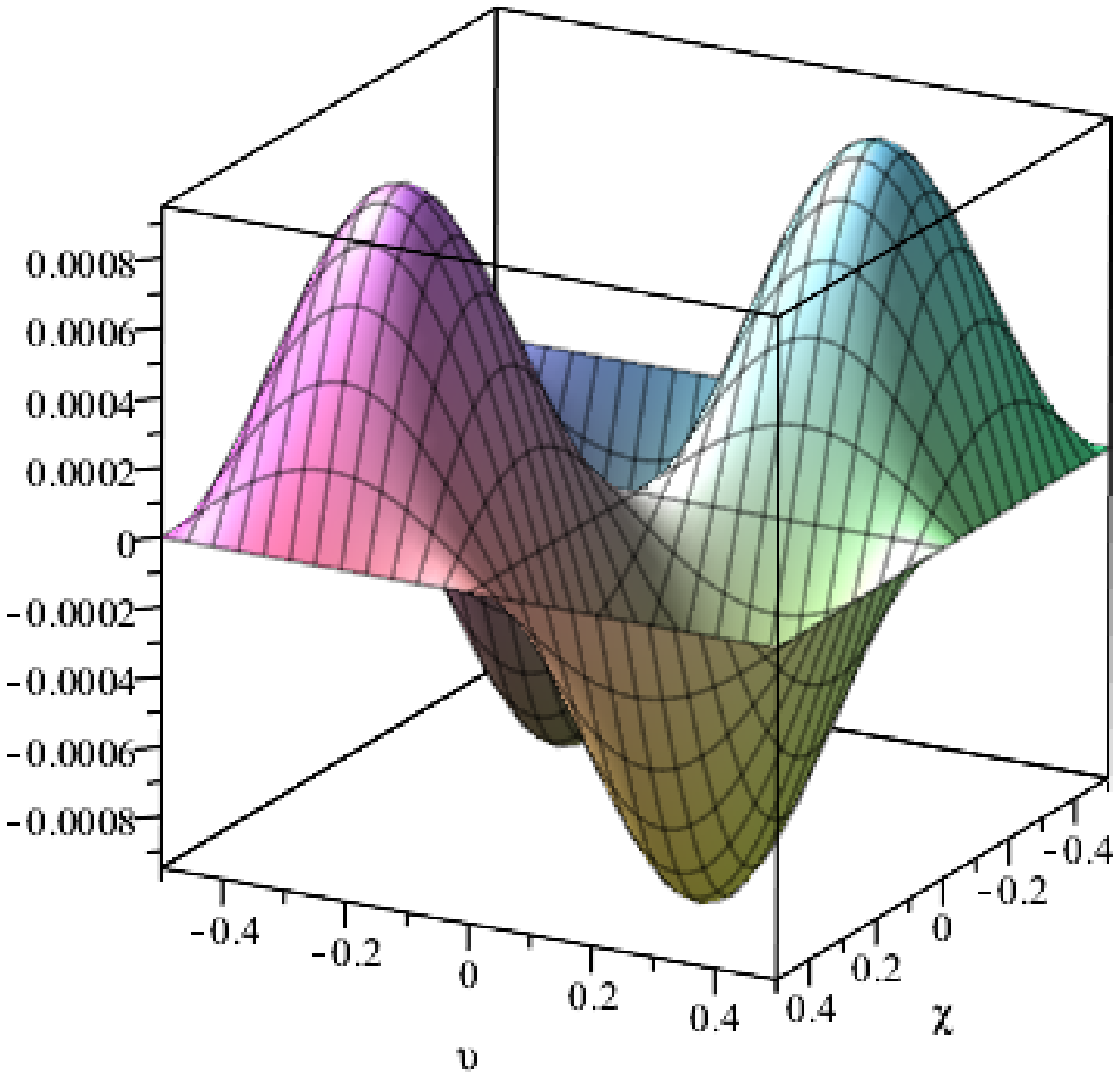} 
\end{centering}
\caption{Ratio $\lambda W^{(1)}/|W^{(0)}|$,
for the interface described by Eq. (\ref{h-sin-sin}), with $L_x=L_y=d=300nm$, 
$\epsilon_1=4$, $\epsilon_2=1$ and $\lambda=1/10$.
The vertical axis exhibits $\lambda W^{(1)}/|W^{(0)}|$, 
whereas the other axes represent $\chi=x/d$ and $\upsilon=y/d$.}
\label{case-sin-sin-W1-W0} 
\end{figure}

\begin{figure}[t]
\begin{centering}
\includegraphics[width=0.8\columnwidth]{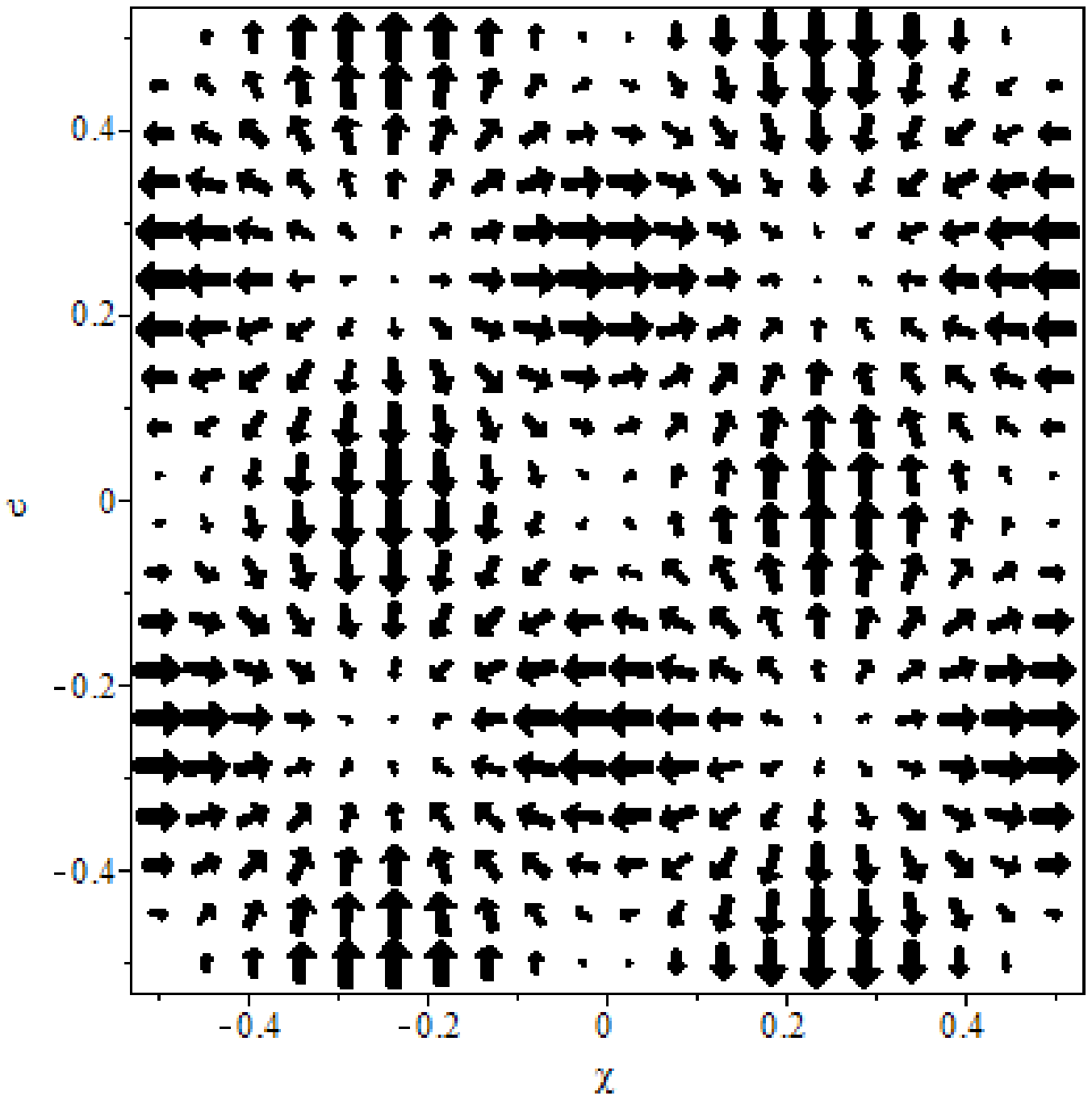} 
\end{centering}
\caption{Effective external force acting on an electron in the two-dimensional system,
for the interface described by Eq. (\ref{h-sin-sin}), with $L_x=L_y=d=300nm$, $\epsilon_1=4$, $\epsilon_2=1$ 
and $\lambda=1/10$. The vectors represent exhibits ${\bf F}_{||}^{(ext)}/|{\bf F}_{\bot}^{(0)}|$,
whereas the axes represent
$\chi=x/d$ and $\upsilon=y/d$.}
\label{case-sin-sin-force-on-an-electron} 
\end{figure}

For the case of a two-dimensional sine grating (see Fig. \ref{case-sin-sin-plane-system-nonplanar-interface}),
\begin{equation}
z= \lambda h\left({\bf r}_{||}\right)=\lambda d\sin\left(k_{x}x\right)\sin\left(k_{y}y\right),
\label{h-sin-sin}
\end{equation}
where and hereafter $k_{x}=2\pi/L_{x}$ and $k_{y}=2\pi/L_{y}$,
we have
\begin{multline}
W^{(1)}=
Q^{2}d\frac{\epsilon_{21}^{-}}{\epsilon_{2}{(\epsilon_{12}^{+})}^{2}}\sin\left(k_{y}y^{\prime}\right)\sin\left(k_{x}x^{\prime}\right)\left[k_{x}^{2}+k_{y}^{2}\right]
\\
\int_{0}^{\infty}d\tilde{R}\tilde{R}\left[\frac{\epsilon_{2}\tilde{R}^{2}+\epsilon_{1}\left[k_{x}^{2}+k_{y}^{2}\right]d^{2}}{\left(\tilde{R}^{2}+\left[k_{x}^{2}+k_{y}^{2}\right]d^{2}\right)^{3}}\right]J_{0}\left(\tilde{R}\right)
\label{W1-sin-sin}
\end{multline}
(whose behavior can be visualized in Fig. \ref{case-sin-sin-W1-W0}),
which is related to the charge-polarized matter interaction
and with an effective external force parallel to the 2D-material, as shown in Fig. \ref{case-sin-sin-force-on-an-electron}.
Note that, since $\epsilon_2<\epsilon_1$, the force points to the next peak of the nonplanar surface.

%%%%%%%%%%%%%%%%%%%%%%%%%%%%%%%%%%%%%%%%%%%%%%%%%%%%%%%%%%%%%%%%%%%%%%%%%%%%%%%%%%%%%%%%%%%%%%%%
\subsection{1D sine-grating}
\label{1D-sine-grating}

\begin{figure}[t]
\begin{centering}
\includegraphics[width=0.8\columnwidth]{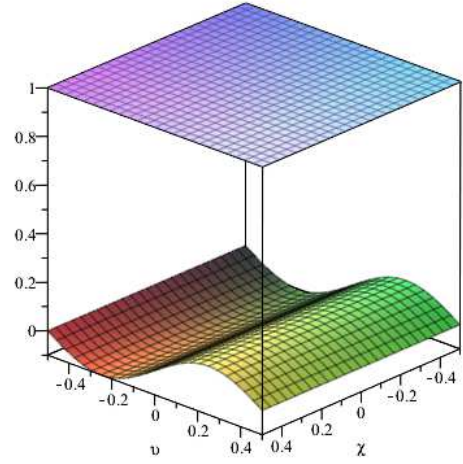} 
\end{centering}
\caption{Illustration of the planar two-dimensional system in the presence
of a nonplanar surface described by Eq. (\ref{h-sin}), with $L_y=d=300nm$,
and $\lambda=1/10$. The vertical axis exhibits $z/d$, whereas the other axis represent
$\chi=x/d$ and $\upsilon=y/d$.}
\label{case-sin-plane-system-nonplanar-interface} 
\end{figure}

\begin{figure}[t]
\begin{centering}
\includegraphics[width=0.8\columnwidth]{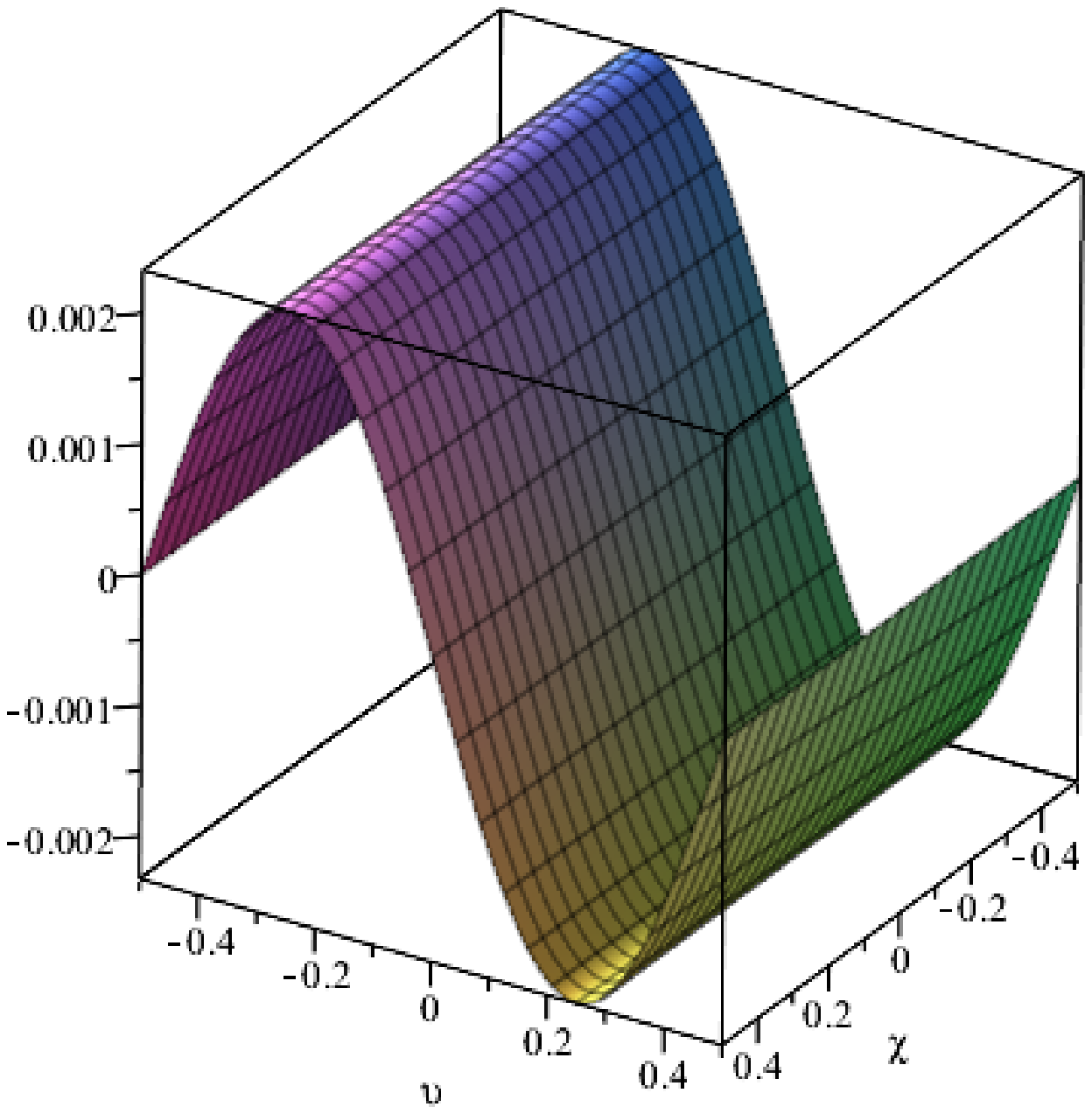} 
\end{centering}
\caption{Ratio $\lambda W^{(1)}/|W^{(0)}|$,
for the interface described by Eq. (\ref{h-sin}), with $L_y=d=300nm$, 
$\epsilon_1=1$, $\epsilon_2=4$ and $\lambda=1/10$.
The vertical axis exhibits $\lambda W^{(1)}/|W^{(0)}|$, 
whereas the other axes represent $\chi=x/d$ and $\upsilon=y/d$.}
\label{case-sin-W1-W0} 
\end{figure}

\begin{figure}[t]
\begin{centering}
\includegraphics[width=0.8\columnwidth]{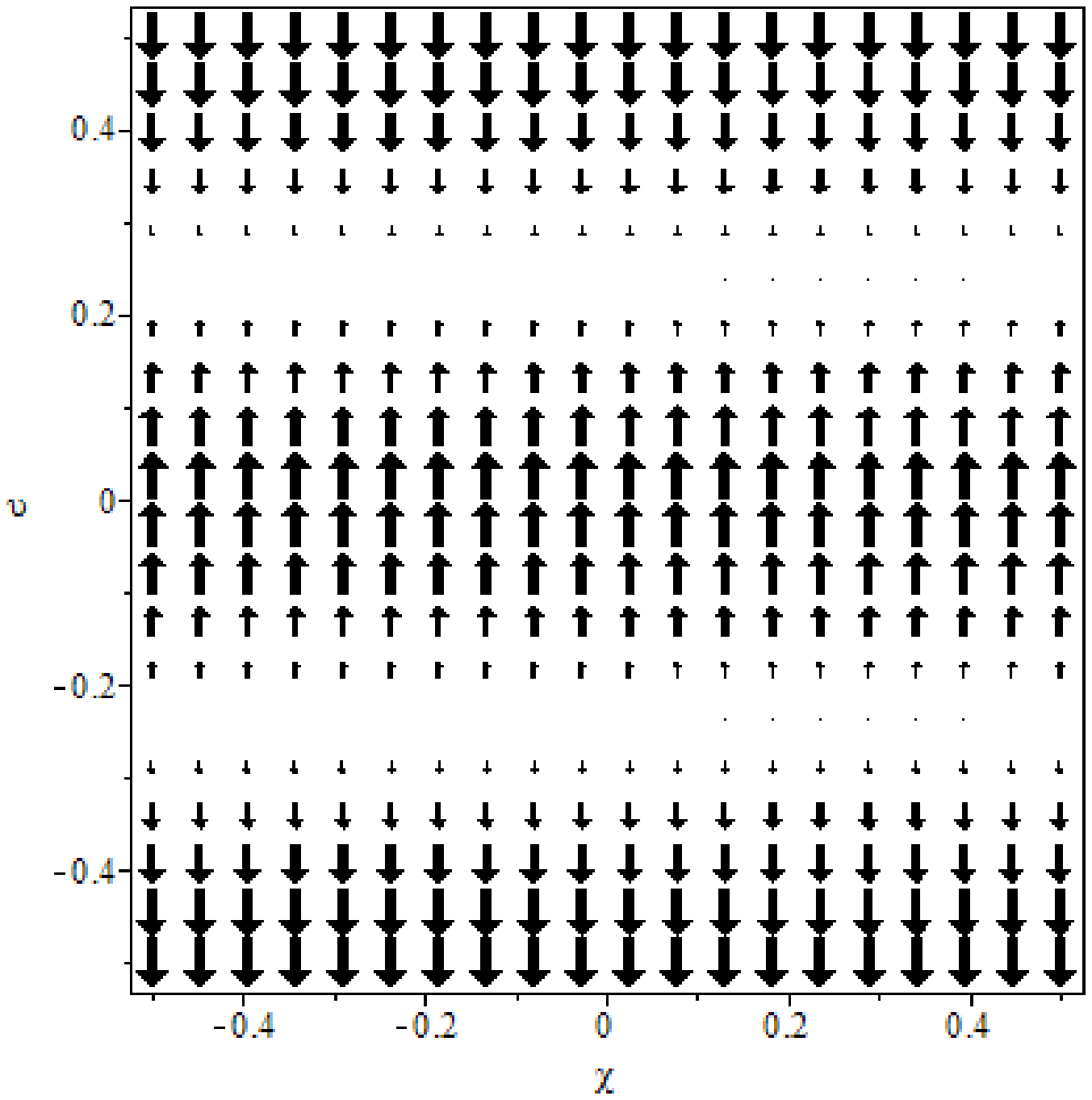} 
\end{centering}
\caption{Effective external force acting on an electron in the two-dimensional system,
for the interface described by Eq. (\ref{h-sin}), with $L_y=d=300nm$, $\epsilon_1=1$, $\epsilon_2=4$ 
and $\lambda=1/10$. The vectors represent exhibits ${\bf F}_{||}^{(ext)}/|{\bf F}_{\bot}^{(0)}|$,
whereas the axes represent $\chi=x/d$ and $\upsilon=y/d$.}
\label{case-sin-force-on-an-electron} 
\end{figure}

For the case of a one-dimensional sine grating (see Fig. \ref{case-sin-plane-system-nonplanar-interface})
\begin{equation}
z= \lambda h\left({\bf r}_{||}\right)=\lambda d\sin\left(k_{y}y\right),
\label{h-sin}
\end{equation}
with $\epsilon_3=\epsilon_2$, we have
\begin{eqnarray}
W^{(1)}&=&Q^{2}d\frac{\epsilon_{21}^{-}}{\epsilon_{2}{(\epsilon_{12}^{+})}^{2}}\sin\left(k_{y}y^{\prime}\right)k_{y}^{2}
\\
&&\int_{0}^{\infty}d\tilde{R}\tilde{R}\left[\frac{\epsilon_{2}\tilde{R}^{2}+\epsilon_{1}k_{y}^{2}d^{2}}{\left(\tilde{R}^{2}+k_{y}^{2}d^{2}\right)^{3}}\right]J_{0}\left(\tilde{R}\right)
\end{eqnarray}
(whose behavior can be visualized in Fig. \ref{case-sin-W1-W0}),
which is related to the interaction charge-polarized matter 
and the effective external force shown in Fig. \ref{case-sin-force-on-an-electron}.
Note that, since $\epsilon_2>\epsilon_1$, the force points to the next valley of the nonplanar surface.

The effective potential related to a charge $Q$ is given by Eq. (\ref{phi-eff-0-1}), with
\begin{multline}
\phi^{\left(0\right)}\left({\bf r}_{||},d,{\bf r}_{||}^{\prime},d\right)=Q\frac{1}{\pi}\int_{0}^{\infty}dq\frac{\left[\epsilon_{21}^{-}e^{-q2d}+\epsilon_{12}^{+}\right]}{2\epsilon_{2}\epsilon_{12}^{+}}
\\
J_{0}\left(q|{\bf r}_{||}^{\prime}-{\bf r}_{||}|\right),
\end{multline}
\begin{multline}
\phi^{\left(1\right)}\left({\bf r}_{||},d,{\bf r}_{||}^{\prime},d\right)=\epsilon_{21}^{-}\frac{Q}{4\pi}\frac{d}{\left(\epsilon_{12}^{+}\right)^{2}}\int d\tilde{x}\int d\tilde{y}\sin\left(k_{y}\tilde{y}\right)
\\
\left\{ \frac{\left[\left(\tilde{x}-x\right)\left(\tilde{x}-x^{\prime}\right)+\left(\tilde{y}-y\right)\left(\tilde{y}-y^{\prime}\right)\right]+\frac{\epsilon_{1}}{\epsilon_{2}}d^{2}}{\left(|{\bf \tilde{r}}_{||}-{\bf r}_{||}|^{2}+d^{2}\right)^{3/2}\left(|{\bf \tilde{r}}_{||}-{\bf r}_{||}^{\prime}|^{2}+d^{2}\right)^{3/2}}\right\}. 
\end{multline}
The term $\phi^{\left(0\right)}$ depends on the distance $|{\bf r}_{||}^{\prime}-{\bf r}_{||}|$,
as expected, whereas the first correction $\phi^{\left(1\right)}$ depends on ${\bf r}_{||}^{\prime}$ and ${\bf r}_{||}$,
separately.

%%%%%%%%%%%%%%%%%%%%%%%%%%%%%%%%%%%%%%%%%%%%%%%%%%%%%%%%%%%%%%%%%%%%%%%%%%%%%%%%%%%%%%%%%%%%%%%%
\section{Final comments and implications of the results}
\label{comments-implications}

Two-dimensional materials, for instance graphene and transition metal dichalcogenide monolayers (TMD),
are very important systems in condensed matter physics.
The behavior of 2D systems between substrates, or in the presence of other material media 
(for instance, conducting materials), is a relevant problem, since
these external media affect, for instance, the Coulomb interaction between electrons in the 2D systems which, in turn, 
influences various electronic properties of these systems.

In the present paper, we have extended the perturbative method for solving Poisson's equation for
a point charge in the presence of a nonplanar conducting interface, proposed by Clinton, Esrick and Sacks \cite{Clinton-1985-II},
to the problem of a point charge between two media with different dielectric constants and in 
the presence of a third dielectric medium separated from those by a nonplanar interface.
Up to the first order $\lambda G^{(1)}$,
we obtained the effective potential, effective electrostatic field, dielectric constant,
and  the effective external field acting along the 2D system. 

The results for $G^{(1)}$, from Eq. (\ref{G_{III}-general}) to (\ref{G_{I}-general}), 
generalize those found in the literature for the case of a vacuum-conductor
situation \cite{Clinton-1985-II}. The results for a vacuum-dielectric interface, 
given by Eqs. (\ref{G_{II}-vacuum-dielectric}) and (\ref{G_{I}-vacuum-dielectric}),
also generalize those found in the literature \cite{Clinton-1985-II}. 
Moreover, Eqs. (\ref{G_{II}-vacuum-conductor}) and (\ref{G_{I}-vacuum-conductor})
recover the vacuum-conducting result found in the literature \cite{Clinton-1985-II} which,
in turn, is formally identical to Hadamard's theorem for Green's functions \cite{Clinton-1985-II,Hadamard-1910}.

In the case where all interfaces are flat, $G=G^{(0)}$ and coincides
with the results found in the literature \cite{Nuno-Livro,Katsnelson-2011}. 
The effective potential, dielectric constant, and electric field are given, respectively,
in Eqs. (\ref{phi-eff-fourier}), (\ref{epsilon-eff}), and (\ref{E-eff-0-1}).
The first terms in the right hand sides of these equations correspond to results
found in the literature \cite{Profumo-2010,Katsnelson-2011,Nuno-Livro},
whereas the second terms (proportional to $\lambda$) correspond to 
the first order correction from the nonplanar behavior, obtained here.

From  Eq. (\ref{phi-eff-fourier}), we obtained that the 
effective potential is affected locally (term $\lambda\phi^{(1)}$) by the presence of the nonplanar interface.
This means that, for example, if a graphene sheet is put on the plane
$z/d=1$ (see Figs. \ref{case-gaussian-plane-system-nonplanar-interface},
\ref{case-sin-sin-plane-system-nonplanar-interface}, and
\ref{case-sin-plane-system-nonplanar-interface}), a local change in the electron-electron interaction
caused by the presence of a nonplanar interface implies in a local renormalization of the Fermi velocity, which, in turn,
can lead to a local increasing of the optical conductivity.
From  Eq. (\ref{E-eff-0-1}), we obtained that the 
effective electric field is affected locally by the presence of the nonplanar interface and
does not point along the line from the source charge to the point where the field is considered.

We have shown that on each charge $Q$ in the 2D planar system acts along the plane an effective external force, 
given by Eq. (\ref{F-||-ext}), which depends on the magnitude of the charge (specifically, on $Q^2$) and
whose direction depends on $\epsilon_2-\epsilon_1$. 
%s. 
This force can point to the next peak of the nonplanar interface (if $\epsilon_2<\epsilon_1$, as illustrated in Fig. \ref{case-sin-sin-force-on-an-electron}), or to a valley (if $\epsilon_2>\epsilon_1$, as illustrated in Fig. \ref{case-sin-force-on-an-electron}). 
The possibility of the effective external force moving the charge in the 2D system to a valley or to a peak, depending on $\epsilon_2-\epsilon_1$,
generalizes the result found in the literature for the case vacuum-conductor,
where the charge is always attracted to a position of the plane which is over the next elevated part of 
the interface \cite{Clinton-1985-I}.
This effective external field, induced by a nonplanar interface, can contribute to the redistribution of the charges in the 2D system,
as, for instance, of electrons in a graphene sheet.

Our results are very general and can be applied in a wide range of other problems.
For instance, in the context of the pseudo-quantum electrodynamics (PQED) \cite{marino-1993,Marino-2017-livro},
an effective quantum field theory describing 2D systems in the presence of nonpanar interfaces
(as illustrated in Fig. \ref{3-camadas-1-rugosa-carga}) needs to be built taking into account 
an effective static potential which is not a Coulombian potential, but in the one given by Eq. (\ref{phi-eff-fourier}).
In addition, the effective 2D quantum field theory should take into account the 
presence of an effective external field [see Eq. (\ref{F-||-ext})] induced by the nonplanar interface.
The formulas obtained in the present paper can also be useful, for example, for problems of finding the quantum states of
electrons localized at surfaces of materials which exhibit
negative electron affinity, in realistic contexts, since
the effects of corrugations on the image potential can be relevant
because it is almost impossible to create perfectly
planar interfaces \cite{Maradudin-1980}. 

Finally, we have obtained the first perturbative correction $G^{(1)}$ in Eq. (\ref{G-expanded}), which is the solution of 
Eq. (\ref{gauss-law}) in the presence of three dielectric regions, as presented in Eq. (\ref{epsilon-com-chaves}). 
On the other hand, procedures similar to those given here can be used to extend the calculations to investigate
systems with a larger number of dielectric regions, or to find other orders of corrections,
enhancing the accuracy of the results.

%%%%%%%%%%%%%%%%%%%%%%%%%%%%%%%%%%%%%%%%%%%%%%%%%%%%%%%%%%%%%%%%%%%%%%%%%%%%%%%%%%%%%%%%%%%%%%%%
\appendix
%%%%%%%%%%%%%%%%%%%%%%%%%%%%%%%%%%%%%%%%%%%%%%%%%%
\section{Solutions for $G^{(0)}\left({\bf q},z,{\bf r}_{||}^{\prime},d\right)$ and $G^{(1)}\left({\bf q},z,{\bf r}_{||}^{\prime},d\right)$}
\label{formulas}

When we have a plane interfaces between the media $\epsilon_3$ and $\epsilon_2$ and
between $\epsilon_2$ and $\epsilon_1$, namely
\begin{equation}
\epsilon\left({\bf r}\right)=\epsilon\left(z\right)=\epsilon_{3}\theta\left[z-d\right]+\epsilon_{2}\theta\left[z\right]\theta\left[d-z\right]+\epsilon_{1}\theta\left[-z\right],
\end{equation}
the solution $\phi^{(0)}$ of Eq. (\ref{gauss-law}) [or the solution
of Eq. (\ref{G0-equation})] can be obtained directly via image method or solving directly this 
equation. The correspondent Fourier version  (noting that only ${\bf r}_{||}$ is involved in the Fourier
transform) of Eq. (\ref{gauss-law}) for this case is \cite{Nuno-Livro}
\begin{multline}
q^{2}\epsilon\left(z\right)G^{(0)}\left({\bf q},z,{\bf r}_{||}^{\prime},d\right)-\frac{\partial}{\partial z}\left[\epsilon\left(z\right)\frac{\partial G^{(0)}\left({\bf q},z,{\bf r}_{||}^{\prime},d\right)}{\partial z}\right]
\\
=4\pi\delta\left(z-d\right)e^{-i{\bf q}\cdot{\bf r}_{||}^{\prime}}.
\end{multline}
Integrating this equation in $z$, between $d-\eta$ and $d+\eta$ and sending $\eta\rightarrow0$
we get
\begin{multline}
\;\;\;\; \epsilon_{3}\left[\frac{\partial}{\partial z}G^{(0)}\left({\bf q},z,{\bf r}_{||}^{\prime},d\right)\right]_{z=d+\eta}
\\
-\epsilon_{2}\left[\frac{\partial}{\partial z}G^{(0)}\left({\bf q},z,{\bf r}_{||}^{\prime},d\right)\right]_{z=d-\eta}=-4\pi e^{-i{\bf q}\cdot{\bf r}_{||}^{\prime}}.
\end{multline}
Integrating again, now between $-\eta$ and $+\eta$,
we get
\begin{eqnarray}
\epsilon_{2}\left[\frac{\partial}{\partial z}G^{(0)}\left({\bf q},z,{\bf r}_{||}^{\prime},z^{\prime}\right)\right]_{z=+\eta}
\;\;\;\;\;\;\;
\\
-\epsilon_{1}\left[\frac{\partial}{\partial z}G^{(0)}\left({\bf q},z,{\bf r}_{||}^{\prime},z^{\prime}\right)\right]_{z=-\eta}=0.
\end{eqnarray}
These equations, together with the continuity condition for $G^{(0)}\left({\bf q},z,{\bf r}_{||}^{\prime},z^{\prime}\right)$ 
and the requirement of $\lim_{z\rightarrow \pm \infty}G^{(0)}=0$, lead to solution in the form shown in Eq. (\ref{G0-solution}),
with:
\begin{equation}
G_{III}^{\left(0\right)}\left({\bf q},z,{\bf r}_{||}^{\prime},d\right)=D^{(0)}\left({\bf q},{\bf r}_{||}^{\prime},d\right)		e^{-qz},
\end{equation}
\begin{eqnarray}
G_{II}^{\left(0\right)}\left({\bf q},z,{\bf r}_{||}^{\prime},d\right)&=&B^{(0)}\left({\bf q},{\bf r}_{||}^{\prime},d\right)e^{qz}
\nonumber
\\
&&+C^{(0)}\left({\bf q},{\bf r}_{||}^{\prime},d\right)e^{-qz},
\end{eqnarray}
\begin{equation}
G_{I}^{\left(0\right)}\left({\bf q},z,{\bf r}_{||}^{\prime},d\right)=A^{(0)}\left({\bf q},{\bf r}_{||}^{\prime},d\right)e^  {qz}
\label{GI0}
\end{equation}
\begin{align}
A^{(0)}\left({\bf q},{\bf r}_{||}^{\prime},d\right)=
\frac{\epsilon_{2}8\pi e^{-i{\bf q}\cdot{\bf r}_{||}^{\prime}}}{q\left[\epsilon_{23}^{-}\epsilon_{12}^{-}e^{-qd}+\epsilon_{23}^{+}\epsilon_{12}^{+}e^{qd}\right]}, 
\label{A0}
\end{align}
\begin{align}
B^{(0)}\left({\bf q},{\bf r}_{||}^{\prime},d\right)=
\frac{\epsilon_{12}^{+}4\pi e^{-i{\bf q}\cdot{\bf r}_{||}^{\prime}}}{q\left[\epsilon_{23}^{-}\epsilon_{12}^{-}e^{-qd}+\epsilon_{23}^{+}\epsilon_{12}^{+}e^{qd}\right]},
\end{align}
\begin{align}
C^{(0)}\left({\bf q},{\bf r}_{||}^{\prime},d\right)=
\frac{\epsilon_{21}^{-}4\pi e^{-i{\bf q}\cdot{\bf r}_{||}^{\prime}}}{q\left[\epsilon_{23}^{-}\epsilon_{12}^{-}e^{-qd}+\epsilon_{23}^{+}\epsilon_{12}^{+}e^{qd}\right]},
\end{align}
\begin{align}
D^{(0)}\left({\bf q},{\bf r}_{||}^{\prime},d\right)=
\frac{\left[\epsilon_{21}^{-}e^{-qd}+\epsilon_{12}^{+}e^{qd}\right]4\pi e^{-i{\bf q}\cdot{\bf r}_{||}^{\prime}}}{q\left[\epsilon_{23}^{-}\epsilon_{12}^{-}e^{-2qd}+\epsilon_{23}^{+}\epsilon_{12}^{+}\right]}. 
\end{align}

Now, let us focus on the solution form $G^{(1)}$.
When we have a plane interface between the media $\epsilon_3$ and $\epsilon_2$, but
a nonplanar interface between $\epsilon_2$ and $\epsilon_1$, as described by Eq. (\ref{epsilon-practical}),
we obtain the solution for $\phi$ in Eq. (\ref{gauss-law})
via perturbative method, according to Eq. (\ref{G-expanded}). The first correction to $G^{(0)}$,  
namely $G^{(1)}$, can be obtained by solving Eq. (\ref{G1-equation}) (in coordinate space) or
Eq. (\ref{G1-equation-fourier}) in Fourier space. The procedures to solve this latter equation are
described in Sec. \ref{main-calculations}, with the functions mentioned in Eq. (\ref{G1-solution})
given by:
\begin{equation}
G_{III}^{\left(1\right)}\left({\bf q},z,{\bf r}_{||}^{\prime},d\right)=D^{(1)}\left({\bf q},{\bf r}_{||}^{\prime},d\right)e^{-qz},
\label{Ap-G-III}
\end{equation}
\begin{eqnarray}
G_{II}^{\left(1\right)}\left({\bf q},z,{\bf r}_{||}^{\prime},d\right)
&=&
B^{(1)}\left({\bf q},{\bf r}_{||}^{\prime},d\right)e^{qz}
\nonumber
\\
&&+C^{(1)}\left({\bf q},{\bf r}_{||}^{\prime},d\right)e^{-qz},
\label{Ap-G-II}
\end{eqnarray}
\begin{equation}
G_{I}^{\left(1\right)}\left({\bf q},z,{\bf r}_{||}^{\prime},d\right)=A^{(1)}\left({\bf q},{\bf r}_{||}^{\prime},d\right)e^{qz},
\label{Ap-G-I}
\end{equation}
\begin{align}
A^{(1)}\left({\bf q},{\bf r}_{||}^{\prime},d\right)=
\frac{\epsilon_{23}^{-}\left(T_{2}\epsilon_{2}+T_{1}\epsilon_{2}q-T_{2}\epsilon_{1}\right)e^{-qd}}{q\left[\epsilon_{23}^{-}\epsilon_{12}^{-}e^{-qd}+\epsilon_{23}^{+}\epsilon_{12}^{+}e^{qd}\right]}
\nonumber\\
-\frac{\epsilon_{23}^{+}\left(-T_{2}\epsilon_{2}+T_{1}\epsilon_{2}q+T_{2}\epsilon_{1}\right)e^{qd}}{q\left[\epsilon_{23}^{-}\epsilon_{12}^{-}e^{-qd}+\epsilon_{23}^{+}\epsilon_{12}^{+}e^{qd}\right]},
\label{Ap-A-1}
\end{align}
\begin{align}
B^{(1)}\left({\bf q},{\bf r}_{||}^{\prime},d\right)=
\frac{\epsilon_{23}^{-}\left(-T_{2}\epsilon_{1}+T_{2}\epsilon_{2}+T_{1}\epsilon_{1}q\right)e^{-qd}}{q\left[\epsilon_{23}^{-}\epsilon_{12}^{-}e^{-qd}+\epsilon_{23}^{+}\epsilon_{12}^{+}e^{qd}\right]},
\label{Ap-B-1}
\end{align}
\begin{align}
C^{(1)}\left({\bf q},{\bf r}_{||}^{\prime},d\right)=
\frac{\epsilon_{23}^{+}\left(-T_{2}\epsilon_{1}+T_{2}\epsilon_{2}+T_{1}\epsilon_{1}q\right)e^{qd}}{q\left[\epsilon_{23}^{-}\epsilon_{12}^{-}e^{-qd}+\epsilon_{23}^{+}\epsilon_{12}^{+}e^{qd}\right]},
\label{Ap-C-1}
\end{align}
\begin{align}
D^{(1)}\left({\bf q},{\bf r}_{||}^{\prime},d\right)=
\frac{2\epsilon_{2}\left(-T_{2}\epsilon_{1}+T_{2}\epsilon_{2}+T_{1}\epsilon_{1}q\right)e^{qd}}{q\left[\epsilon_{23}^{-}\epsilon_{12}^{-}e^{-qd}+\epsilon_{23}^{+}\epsilon_{12}^{+}e^{qd}\right]},
\label{Ap-D-1}
\end{align}
with
\begin{align}
T_{1}=&-\frac{\epsilon_{12}^{-}}{\epsilon_{2}}\int\frac{1}{\left(2\pi\right)^{2}}d^{2}{\bf q^{\prime}}h\left({\bf q}-{\bf q^{\prime}}\right)
q^{\prime}G^{\left(0\right)}\left({\bf q}^{\prime},0,{\bf r}_{||}^{\prime},d\right),
\label{Ap-T-1}
\end{align}
\begin{align}
T_{2}=\int\frac{1}{\left(2\pi\right)^{2}}d^{2}{\bf q^{\prime}}h\left({\bf q}-{\bf q^{\prime}}\right)\left({\bf q^{\prime}}\cdot{\bf q}\right)
G^{\left(0\right)}\left({\bf q}^{\prime},0,{\bf r}_{||}^{\prime},d\right),
\label{Ap-T-2}
\end{align}
\begin{align}
G^{\left(0\right)}\left({\bf q}^{\prime},0,{\bf r}_{||}^{\prime},d\right)=
\frac{\epsilon_{2}8\pi e^{-i{\bf q}^{\prime}\cdot{\bf r}_{||}^{\prime}}}{q\left[\epsilon_{23}^{-}\epsilon_{12}^{-}e^{-q^{\prime}d}+\epsilon_{23}^{+}\epsilon_{12}^{+}e^{q^{\prime}d}\right]}.
\label{Ap-G-0}
\end{align}
%

%%%%%%%%%%%%%%%%%%%%%%%%%%%%%%%%%%%%%%%%%%%%%%%%%%
\section{Boundary conditions}
\label{app-bc}

Let us start, considering:
\begin{equation}
 G\left({\bf r}_{||},z,{\bf r}_{||}^{\prime},d\right) = 
  \begin{cases} 
   G_{III}\left({\bf r}_{||},z,{\bf r}_{||}^{\prime},d\right),\; d < z\\
	 G_{II}\left({\bf r}_{||},z,{\bf r}_{||}^{\prime},d\right),\; \lambda h\left({\bf r}_{||}\right)<z<d \\
   G_{I}\left({\bf r}_{||},z,{\bf r}_{||}^{\prime},d\right),\; z<\lambda h\left({\bf r}_{||}\right).
	\label{G-solution-r}
  \end{cases}
\end{equation}
Requiring the continuity of the Green function, we have
\begin{equation}
G_{II}\left({\bf r}_{||},\lambda h\left({\bf r}_{||}\right),{\bf r}_{||}^{\prime},d\right)=G_{I}\left({\bf r}_{||},\lambda h\left({\bf r}_{||}\right),{\bf r}_{||}^{\prime},d\right),
\label{main-condition-h}
\end{equation}
from which we get
\begin{multline}
G_{II}\left({\bf r}_{||},0,{\bf r}_{||}^{\prime},d\right)
+\left[\frac{\partial}{\partial z}G_{II}\left({\bf r}_{||},z,{\bf r}_{||}^{\prime},d\right)\right]_{z=0}\lambda h\left({\bf r}_{||}\right)+...
\\
=G_{I}\left({\bf r}_{||},0,{\bf r}_{||}^{\prime},d\right)+\left[\frac{\partial}{\partial z}G_{I}\left({\bf r}_{||},z,{\bf r}_{||}^{\prime},d\right)\right]_{z=0}\lambda h\left({\bf r}_{||}\right)+...
\label{condition-h-toll}
\end{multline}
Using Eq. (\ref{G-expanded}) in Eq. (\ref{condition-h-toll}), we have
\begin{multline}
G_{II}^{(0)}\left({\bf r}_{||},0,{\bf r}_{||}^{\prime},z^{\prime}\right)+\lambda G_{II}^{(1)}\left({\bf r}_{||},0,{\bf r}_{||}^{\prime},z^{\prime}\right)
\\
+\left[\frac{\partial}{\partial z}G_{II}^{(0)}\left({\bf r}_{||},z,{\bf r}_{||}^{\prime},z^{\prime}\right)\right]_{z=0}\lambda h\left({\bf r}_{||}\right)+\mathcal{O}\left(\lambda^{2}\right)
\\
=G_{I}^{(0)}\left({\bf r}_{||},0,{\bf r}_{||}^{\prime},z^{\prime}\right)+\lambda G_{I}^{(1)}\left({\bf r}_{||},0,{\bf r}_{||}^{\prime},z^{\prime}\right)
\\
+
\left[\frac{\partial}{\partial z}G_{I}^{(0)}\left({\bf r}_{||},z,{\bf r}_{||}^{\prime},z^{\prime}\right)\right]_{z=0}\lambda h\left({\bf r}_{||}\right)+\mathcal{O}\left(\lambda^{2}\right),
\end{multline}
from which we obtain the two boundary conditions written next.
First, for $G^{(0)}$, we have
\begin{equation}
G_{II}^{(0)}\left({\bf {\bf r}_{||}},0,{\bf r}_{||}^{\prime},z^{\prime}\right)=G_{I}^{(0)}\left({\bf {\bf r}_{||}},0,{\bf r}_{||}^{\prime},z^{\prime}\right),
\end{equation}
whose Fourier version is
\begin{equation}
G_{II}^{(0)}\left({\bf q},0,{\bf r}_{||}^{\prime},z^{\prime}\right)=G_{I}^{(0)}\left({\bf q},0,{\bf r}_{||}^{\prime},z^{\prime}\right),
\end{equation}
which we also write as
\begin{equation}
G^{(0)}\left({\bf q},0^{+},{\bf r}_{||}^{\prime},z^{\prime}\right)=G^{(0)}\left({\bf q},0^{-},{\bf r}_{||}^{\prime},z^{\prime}\right),
\end{equation}
used in Eq. (\ref{bc-G1-1}).
Second, for $G^{(1)}$, we obtain
\begin{multline}
G_{II}^{(1)}\left({\bf r}_{||},0,{\bf r}_{||}^{\prime},d\right)-G_{I}^{(1)}\left({\bf r}_{||},0,{\bf r}_{||}^{\prime},d\right)
\\	
=-\bigg\{ \left[\frac{\partial}{\partial z}G_{II}^{(0)}\left({\bf r}_{||},z,{\bf r}_{||}^{\prime},d\right)\right]_{z=0}
\\
-\left[\frac{\partial}{\partial z}G_{I}^{(0)}\left({\bf r}_{||},z,{\bf r}_{||}^{\prime},d\right)\right]_{z=0}\bigg\} h\left({\bf r}_{||}\right),
\end{multline}
which can be written as
\begin{multline}
G^{(1)}\left({\bf r}_{||},0^{+},{\bf r}_{||}^{\prime},d\right)-G^{(1)}\left({\bf r}_{||},0^{-},{\bf r}_{||}^{\prime},d\right)
\\	
=-\bigg\{ \left[\frac{\partial}{\partial z}G^{(0)}\left({\bf r}_{||},z,{\bf r}_{||}^{\prime},d\right)\right]_{z=0^{+}}
\\
-\left[\frac{\partial}{\partial z}G^{(0)}\left({\bf r}_{||},z,{\bf r}_{||}^{\prime},d\right)\right]_{z=0^{-}}\bigg\} h\left({\bf r}_{||}\right),
\end{multline}
whose Fourier version is shown in Eq. (\ref{bc-G1-2}).

For the region $z=d$, we require the following continuity condition
for the Green function:
\begin{equation}
G_{III}\left({\bf q},d,{\bf r}_{||}^{\prime},z^{\prime}\right)=G_{II}\left({\bf q},d,{\bf r}_{||}^{\prime},z^{\prime}\right). 
\label{main-condition-d}
\end{equation}
Expanding this equation, we have
\begin{multline}
G_{III}^{(0)}\left({\bf q},d,{\bf r}_{||}^{\prime},z^{\prime}\right)+\lambda G_{III}^{(1)}\left({\bf q},d,{\bf r}_{||}^{\prime},z^{\prime}\right)+\mathcal{O}\left(\lambda^{2}\right)
\\
=G_{II}^{(0)}\left({\bf q},d,{\bf r}_{||}^{\prime},z^{\prime}\right)+\lambda G_{II}^{(1)}\left({\bf q},d,{\bf r}_{||}^{\prime},z^{\prime}\right)+\mathcal{O}\left(\lambda^{2}\right),
\end{multline}
from which we obtain the other two boundary conditions.
For $G^{(0)}$, we get
\begin{equation}
G_{III}^{(0)}\left({\bf q},d,{\bf r}_{||}^{\prime},z^{\prime}\right)=G_{II}^{(0)}\left({\bf q},d,{\bf r}_{||}^{\prime},z^{\prime}\right),
\end{equation}
which can be written in the notation
\begin{equation}
G^{(0)}\left({\bf q},d^{+},{\bf r}_{||}^{\prime},z^{\prime}\right)=G^{(0)}\left({\bf q},d^{-},{\bf r}_{||}^{\prime},z^{\prime}\right).
\end{equation}
For $G^{(1)}$ we get
\begin{equation}
G_{III}^{(1)}\left({\bf q},d,{\bf r}_{||}^{\prime},z^{\prime}\right)=G_{II}^{(1)}\left({\bf q},d,{\bf r}_{||}^{\prime},z^{\prime}\right),
\end{equation}
which can be written in the notation 
\begin{equation}
G^{(1)}\left({\bf q},d^{+},{\bf r}_{||}^{\prime},z^{\prime}\right)=G^{(1)}\left({\bf q},d^{-},{\bf r}_{||}^{\prime},z^{\prime}\right),
\end{equation}
used in Eq. (\ref{bc-G1-1}).

%%%%%%%%%%%%%%%%%%%%%%%%%%%%%%%%%%%%%%%%%%%%%%%%%%
\section{Obtaining $G^{\left(1\right)}\left({\bf r}_{||},z,{\bf r}_{||}^{\prime},d\right)$}
\label{G-1-x-y}

The solution for $G^{\left(1\right)}$ in terms of  $x$ and $y$ can be 
obtained using Eqs. (\ref{fourier-representation}), (\ref{G1-solution}),
and (\ref{Ap-G-III}) - (\ref{Ap-G-0}). This leads to: 
\begin{multline}
G_{III}^{\left(1\right)}\left({\bf r}_{||},z,{\bf r}_{||}^{\prime},z^{\prime}\right)=\epsilon_{21}^{-}\frac{1}{4\pi}\int\frac{1}{\left(2\pi\right)^{2}}d^{2}{\bf q}\int\frac{1}{\left(2\pi\right)^{2}}d^{2}{\bf q^{\prime}}
\\
h\left({\bf q}-{\bf q^{\prime}}\right)
\Bigg\{{\bf \boldsymbol{\nabla}_{||}}G_{I}^{\left(0\right)}\left({\bf q},d-z,-{\bf r}_{||},d\right)\cdot
\\
{\bf \boldsymbol{\nabla}_{||}^{\prime}}G_{I}^{\left(0\right)}\left({\bf q}^{\prime},0,{\bf r}_{||}^{\prime},z^{\prime}\right)
+\frac{\epsilon_{1}}{\epsilon_{2}}\left[\frac{\partial}{\partial\tilde{z}}G_{I}^{\left(0\right)}\left({\bf q},\tilde{z},-{\bf r}_{||},z^{\prime}\right)\right]_{\tilde{z}=d-z}
\\
\left[\frac{\partial}{\partial\tilde{z}}G_{I}^{\left(0\right)}\left({\bf q}^{\prime},\tilde{z},{\bf r}_{||}^{\prime},z^{\prime}\right)\right]_{\tilde{z}=0}
\Bigg\},\;\;\;\;\;\;\;\;\;\;
\end{multline}
\begin{multline}
G_{II}^{\left(1\right)}\left({\bf r}_{||},z,{\bf r}_{||}^{\prime},z^{\prime}\right)=
\frac{1}{8\pi}\frac{1}{\epsilon_{2}}\epsilon_{21}^{-}\int\frac{1}{\left(2\pi\right)^{2}}d^{2}{\bf q}\int\frac{1}{\left(2\pi\right)^{2}}d^{2}{\bf q^{\prime}}
\\
h\left({\bf q}-{\bf q^{\prime}}\right)
\Bigg\{
\epsilon_{23}^{-}\Bigg[
\left[{\bf \boldsymbol{\nabla}_{||}}G_{I}^{\left(0\right)}\left({\bf q},\tilde{z},-{\bf r}_{||},d\right)\right]_{\tilde{z}=z-d}
\cdot
\\
{\bf \boldsymbol{\nabla}_{||}^{\prime}}G_{I}^{\left(0\right)}\left({\bf q}^{\prime},0,{\bf r}_{||}^{\prime},z^{\prime}\right)
+\frac{\epsilon_{1}}{\epsilon_{2}}\left[\frac{\partial}{\partial\tilde{z}}G_{I}^{\left(0\right)}\left({\bf q},\tilde{z},-{\bf r}_{||},d\right)\right]_{\tilde{z}=z-d}
\\
\left[\frac{\partial}{\partial\tilde{z}}G_{I}^{\left(0\right)}\left({\bf q}^{\prime},\tilde{z},{\bf r}_{||}^{\prime},z^{\prime}\right)\right]_{\tilde{z}=0}
\Bigg]
+
\epsilon_{23}^{+}
\Bigg[
\\
\left[{\bf \boldsymbol{\nabla}_{||}}G_{I}^{\left(0\right)}\left({\bf q},\tilde{z},-{\bf r}_{||},d\right)\right]_{\tilde{z}=d-z}\cdot
{\bf \boldsymbol{\nabla}_{||}^{\prime}}G_{I}^{\left(0\right)}\left({\bf q}^{\prime},0,{\bf r}_{||}^{\prime},z^{\prime}\right)
\\
+\frac{\epsilon_{1}}{\epsilon_{2}}\left[\frac{\partial}{\partial\tilde{z}}G_{I}^{\left(0\right)}\left({\bf q},\tilde{z},-{\bf r}_{||},d\right)\right]_{\tilde{z}=d-z}
\\
\left[\frac{\partial}{\partial\tilde{z}}G_{I}^{\left(0\right)}\left({\bf q}^{\prime},\tilde{z},{\bf r}_{||}^{\prime},z^{\prime}\right)\right]_{\tilde{z}=0}
\Bigg]\Bigg\},\;\;\;\;\;\;\;\;\;\;
\end{multline}
\begin{multline}
G_{I}^{\left(1\right)}\left({\bf r}_{||},z,{\bf r}_{||}^{\prime},d\right)=\frac{1}{8\pi}\frac{\epsilon_{21}^{-}}{\epsilon_{2}}\int\frac{1}{\left(2\pi\right)^{2}}d^{2}{\bf q}
\int\frac{1}{\left(2\pi\right)^{2}}d^{2}{\bf q^{\prime}}
\\
h\left({\bf q}-{\bf q^{\prime}}\right)
\Bigg\{
\epsilon_{23}^{-}\Bigg[{\bf \boldsymbol{\nabla}_{||}}G_{I}^{\left(0\right)}\left({\bf q},z-d,-{\bf r}_{||},d\right)
\cdot
\\
{\bf \boldsymbol{\nabla}_{||}^{\prime}}G_{I}^{\left(0\right)}\left({\bf q}^{\prime},0,{\bf r}_{||}^{\prime},d\right)
+
\left[\frac{\partial}{\partial\tilde{z}}G_{I}^{\left(0\right)}\left({\bf q},\tilde{z},-{\bf r}_{||},d\right)\right]_{\tilde{z}=z-d}
\\
\left[\frac{\partial}{\partial\tilde{z}}G_{I}^{\left(0\right)}\left({\bf q}^{\prime},\tilde{z},{\bf r}_{||}^{\prime},d\right)\right]_{\tilde{z}=0}
\Bigg]
+\epsilon_{23}^{+}
\\
\Bigg[{\bf \boldsymbol{\nabla}_{||}}G_{I}^{\left(0\right)}\left({\bf q},z+d,-{\bf r}_{||},d\right)\cdot
{\bf \boldsymbol{\nabla}_{||}^{\prime}}G_{I}^{\left(0\right)}\left({\bf q}^{\prime},0,{\bf r}_{||}^{\prime},d\right)
-
\\
\left[\frac{\partial}{\partial\tilde{z}}G_{I}^{\left(0\right)}\left({\bf q},\tilde{z},-{\bf r}_{||},d\right)\right]_{\tilde{z}=z+d}
\\
\left[\frac{\partial}{\partial\tilde{z}}G_{I}^{\left(0\right)}\left({\bf q}^{\prime},\tilde{z},{\bf r}_{||}^{\prime},d\right)\right]_{\tilde{z}=0}
\Bigg]\Bigg\}.\;\;\;\;\;\;\;\;\;\;
\end{multline}
Now, considering the symmetry
\begin{equation}
G_{I}^{\left(0\right)}\left({\bf r}_{||},z,{\bf r}_{||}^{\prime},z^{\prime}\right)=G_{I}^{\left(0\right)}\left({\bf -r}_{||},z,-{\bf r}_{||}^{\prime},z^{\prime}\right),
\end{equation}
we get, after manipulations, the formulas (\ref{G_{III}-general})-(\ref{G2}).
%
%%%%%%%%%%%%%%%%%%%%%%%%%%%%%%%%%%%%%%%%%%%%%%%%%%
\section{Energy of interaction}
\label{app-W}

When we put together a set of macroscopic (real) charges 
[described by $\rho\left({\bf r}\right)$], 
in the presence of dielectric media, we have to take into account 
the state of polarization induced in these media \cite{Jackson-livro}.
The total work $W$ to assemble the system described by $\rho\left({\bf r}\right)$ includes the
work done on the dielectric media. 
If the behavior of the media is linear, then we can use the formula \cite{Jackson-livro}
\begin{equation}
W=\frac{1}{2}\int d^{3}r\rho\left({\bf r}\right)\phi\left({\bf r}\right).
\end{equation}
Let us consider the total potential $\phi$ divided into two parts:
\begin{equation}
\phi\left({\bf r}\right)=\phi_{\rho}\left({\bf r}\right)+\phi_{ind}\left({\bf r}\right),
\end{equation}
where $\phi_{\rho}$ is the potential associated with the distribution $\rho\left({\bf r}\right)$,
whereas $\phi_{ind}$ is the potential produced by the averaged induced charges on the dielectric media.
Considering
\begin{equation}
\rho\left({\bf r}\right)=Q\delta\left({\bf r-}{\bf r}^{\prime}\right),
\end{equation}
and using the notation $\phi_{\rho}\rightarrow \phi_{Q}$, we have
\begin{equation}
W=\frac{1}{2}Q\phi_{Q}\left({\bf {\bf r}^{\prime}}\right)+\frac{1}{2}Q\phi_{ind}\left({\bf r}^{\prime}\right).
\label{W-app-Q-ind}
\end{equation}
The term $\frac{1}{2}Q\phi_{Q}\left({\bf {\bf r}^{\prime}}\right)$ can be seen as 
the work to build the point charge $Q$, which is divergent and will be discarded.
Then, effectively, we will consider just the second term in the right hand side of Eq. (\ref{W-app-Q-ind}),
which leads to Eq. (\ref{W-phi-ind}).

%%%%%%%%%%%%%%%%%%%%%%%%%%%%%%%%%%%%%%%%%%%%%%%%%%
\section{Potential and electric fields in coordinate representation}
\label{phi-E-coordinate}

The functions $\phi^{(0)}$ and $\phi^{(1)}$ in Eq. (\ref{phi-eff-0-1}) are given
explicitly by:
\begin{eqnarray}
\phi^{\left(0\right)}\left({\bf r}_{||},d,{\bf r}_{||}^{\prime},d\right)&=&\frac{Q}{\pi}\int_{0}^{\infty}dq\frac{\left[\epsilon_{21}^{-}e^{-q2d}+\epsilon_{12}^{+}\right]}{\left[\epsilon_{23}^{-}\epsilon_{12}^{-}e^{-2qd}+\epsilon_{23}^{+}\epsilon_{12}^{+}\right]}
\nonumber
\\
&&J_{0}\left(q|{\bf r}_{||}^{\prime}-{\bf r}_{||}|\right), 
\label{phi-0-eff}
\end{eqnarray}
\newpage
\begin{multline}
\phi^{\left(1\right)}\left({\bf r}_{||},d,{\bf r}_{||}^{\prime},d\right)=\epsilon_{21}^{-}\epsilon_{2}^{2}\frac{Q}{\pi}\int d^{2}{\bf \tilde{r}}h\left(\tilde{{\bf r}}_{||}\right) 
\\
\int_{0}^{\infty}d\tilde{q}\frac{\tilde{q}}{\left[\epsilon_{23}^{-}\epsilon_{12}^{-}e^{-\tilde{q}d}+\epsilon_{23}^{+}\epsilon_{12}^{+}e^{\tilde{q}d}\right]}
\\
\int_{0}^{\infty}dq\frac{q}{\left[\epsilon_{23}^{-}\epsilon_{12}^{-}e^{-qd}+\epsilon_{23}^{+}\epsilon_{12}^{+}e^{qd}\right]}
\\
\bigg\{
J_{1}\left(\tilde{q}|{\bf \tilde{r}}_{||}-{\bf r}_{||}|\right)J_{1}\left(q|{\bf \tilde{r}}_{||}-{\bf r}_{||}^{\prime}|\right)\mathcal{A}^{\left(1\right)}\left({\bf \tilde{r}}_{||},{\bf r}_{||},{\bf r}_{||}^{\prime}\right)
\\
+\frac{\epsilon_{1}}{\epsilon_{2}}J_{0}\left(\tilde{q}|{\bf \tilde{r}}_{||}-{\bf r}_{||}|\right)J_{0}\left(q|{\bf \tilde{r}}_{||}-{\bf r}_{||}^{\prime}|\right)
\bigg\}
\label{phi-1-eff}
\end{multline}
where
\begin{eqnarray}
\mathcal{A}^{\left(1\right)}\left({\bf \tilde{r}}_{||},{\bf r}_{||},{\bf r}_{||}^{\prime}\right)
&=&
\frac{\left(\tilde{x}-x\right)}{|{\bf \tilde{r}}_{||}-{\bf r}_{||}|}\frac{\left(\tilde{x}-x^{\prime}\right)}{|{\bf \tilde{r}}_{||}-{\bf r}_{||}^{\prime}|}
\\
&&+\frac{\left(\tilde{y}-y\right)}{|{\bf \tilde{r}}_{||}-{\bf r}_{||}|}\frac{\left(\tilde{y}-y^{\prime}\right)}{|{\bf \tilde{r}}_{||}-{\bf r}_{||}^{\prime}|}.
\end{eqnarray}

The fields ${\bf E}_{||}^{\left(0\right)}$ and ${\bf E}_{||}^{\left(1\right)}$ are given by:
\begin{eqnarray}
{\bf E}_{||}^{\left(0\right)}\left({\bf r}_{||},d,{\bf r}_{||}^{\prime},d\right)
&=&
-Q\frac{1}{\pi}\int_{0}^{\infty}dq\frac{\left[\epsilon_{21}^{-}e^{-q2d}+\epsilon_{12}^{+}\right]}{\left[\epsilon_{23}^{-}\epsilon_{12}^{-}e^{-2qd}+\epsilon_{23}^{+}\epsilon_{12}^{+}\right]}
\nonumber
\\
&&
qJ_{1}\left(q|{\bf r}_{||}^{\prime}-{\bf r}_{||}|\right)\frac{{\bf r}_{||}^{\prime}-{\bf r}_{||}}{|{\bf r}_{||}^{\prime}-{\bf r}_{||}|},
\label{E-0-eff}
\end{eqnarray}
\begin{multline}
{\bf E}_{||}^{\left(1\right)}\left({\bf r}_{||},d,{\bf r}_{||}^{\prime},d\right)=\epsilon_{21}^{-}\epsilon_{2}^{2}\frac{4}{\pi}Q\int d^{2}{\bf \tilde{r}}h\left(\tilde{{\bf r}}_{||}\right)
\\
\int_{0}^{\infty}d\tilde{q}\frac{\tilde{q}}{\left[\epsilon_{23}^{-}\epsilon_{12}^{-}e^{-\tilde{q}d}+\epsilon_{23}^{+}\epsilon_{12}^{+}e^{\tilde{q}d}\right]}
\\
\int_{0}^{\infty}dq\frac{q}{\left[\epsilon_{23}^{-}\epsilon_{12}^{-}e^{-qd}+\epsilon_{23}^{+}\epsilon_{12}^{+}e^{qd}\right]}
\\
\bigg\{
\tilde{q}J_{0}\left(\tilde{q}|\boldsymbol{\beta}_{1}|\right)J_{1}\left(q|\boldsymbol{\beta}_{2}|\right)
\mathcal{A}^{\left(1\right)}\left({\bf \tilde{r}}_{||},{\bf r}_{||},{\bf r}_{||}^{\prime},\right)\frac{\boldsymbol{\beta}_{1}}{|\boldsymbol{\beta}_{1}|}
\\
-J_{1}\left(\tilde{q}|\boldsymbol{\beta}_{1}|\right)J_{1}\left(q|\boldsymbol{\beta}_{2}|\right)
\mathcal{A}^{\left(1\right)}\left({\bf \tilde{r}}_{||},{\bf r}_{||},{\bf r}_{||}^{\prime},\right)\frac{\boldsymbol{\beta}_{1}}{|\boldsymbol{\beta}_{1}|^{2}}
\\
+J_{1}\left(\tilde{q}|\boldsymbol{\beta}_{1}|\right)J_{1}\left(q|\boldsymbol{\beta}_{2}|\right)\mathcal{A}^{\left(2\right)}\left({\bf \tilde{r}}_{||},{\bf r}_{||},{\bf r}_{||}^{\prime},\right)\boldsymbol{\beta}_{3}
\\
-\frac{\epsilon_{1}}{\epsilon_{2}}\tilde{q}J_{1}\left(\tilde{q}|\boldsymbol{\beta}_{1}|\right)J_{0}\left(q|\boldsymbol{\beta}_{2}|\right)\frac{\boldsymbol{\beta}_{1}}{|\boldsymbol{\beta}_{1}|}
\bigg\},
\label{E-1-eff}
\end{multline}
where
\begin{eqnarray}
\boldsymbol{\beta}_{1}&=&{\bf \tilde{r}}_{||}-{\bf r}_{||}, 
\\\cr
\boldsymbol{\beta}_{2}&=&{\bf \tilde{r}}_{||}-{\bf r}_{||}^{\prime}, 
\\\cr
\boldsymbol{\beta}_{3}&=&-\left(\tilde{y}-y\right)\hat{{\bf x}}+\left(\tilde{x}-x\right)\hat{{\bf y}},
\\\cr
\mathcal{A}^{\left(2\right)}\left({\bf \tilde{r}}_{||},{\bf r}_{||},{\bf r}_{||}^{\prime}\right)&=&\frac{\left(\tilde{x}-x^{\prime}\right)}{|{\bf \tilde{r}}_{||}-{\bf r}_{||}^{\prime}|}\frac{\left(\tilde{y}-y\right)}{|{\bf \tilde{r}}_{||}-{\bf r}_{||}|^{3}}
\nonumber
\\
&&-\frac{\left(\tilde{y}-y^{\prime}\right)}{|{\bf \tilde{r}}_{||}-{\bf r}_{||}^{\prime}|}\frac{\left(\tilde{x}-x\right)}{|{\bf \tilde{r}}_{||}-{\bf r}_{||}|^{3}}.
\end{eqnarray}

%%%%%%%%%%%%%%%%%%%%%%%%%%%%%%%%%%%%%%%%%%%%%%%%%%%%%%%%%%%%%%%%%%%%%%%%%%%%%%%%%%%%%%%%%%%
\begin{acknowledgments}

\end{acknowledgments}

D.T.A. was partially supported by Coordena\c{c}\~{a}o de Aperfei\c{c}oamento
de Pessoal de N\'{i}vel Superior (CAPES/Brazil), via Programa Est\'{a}gio S\^{e}nior no Exterior -- Processo 88881.119705/2016-01 --, 
by Conselho Nacional de Desenvolvimento Cient\'{i}fico e Tecnol\'{o}gico (CNPq/Brazil), via Processo  461826/2014-3 (Edital Universal),
and also thanks the hospitality of the Centro de F\'{i}sica, Universidade do Minho, Braga -- Portugal.
N.M.R.P. acknowledges support from the European Commission through the project ``Graphene Driven Revolutions in ICT and Beyond'' (Ref. No. 785219), FEDER, and the Portuguese Foundation for Science and Technology (FCT) through project POCI-01-0145-FEDER-028114,
and in the framework of the Strategic Financing UID/FIS/04650/2013.
%%%%%%%%%%%%%%%%%%%%%%%%%%%%%%%%%%%%%%%%%%%%%%%%%%%%%%%%%%%%%%%%%%%%%%%%%%%
%%%%%%%%%%%%%%%%%%%%%%%%%%%%%%%%%%%%%%%%%%%%%%%%%%%%%%%%%%%%%%%%%%%%%%%%%%%

%\bibliography{ourrefs}


\end{document}